\RequirePackage{fix-cm}
\documentclass[twocolumn]{svjour3}          
\smartqed  
\usepackage{graphicx}
\usepackage[square,sort&compress,comma,numbers]{natbib}
\usepackage{microtype}
\usepackage{graphicx}
\usepackage{subfigure}
\usepackage{booktabs} 

\usepackage{datetime}
\usepackage{url}
\usepackage[ruled, noend]{algorithm2e}

\usepackage{float}
\usepackage{color}
\usepackage{multirow}
\usepackage{soul}
\usepackage{array}
\usepackage{subcaption}
\usepackage{soul}
\usepackage{balance}
\usepackage{enumitem}
\usepackage{graphicx}
\usepackage{amsmath}
\usepackage{kotex}
\usepackage{mathtools}
\usepackage{physics}

\usepackage{stmaryrd}
\usepackage{trimclip}
\usepackage{xcolor}
\usepackage{fontawesome5}
\usepackage{needspace}

\usepackage{amsfonts}

\usepackage{pifont}

\makeatletter
\DeclareRobustCommand{\shortto}{%
  \mathrel{\mathpalette\short@to\relax}%
}

\newcommand{\short@to}[2]{%
  \mkern2mu
  \clipbox{{.3\width} 0 0 0}{$\m@th#1\vphantom{+}{\shortrightarrow}$}%
  }
\makeatother

\newcolumntype{L}[1]{>{\raggedright\let\newline\\\arraybackslash\hspace{0pt}}m{#1}}

\usepackage{tikz}
\usetikzlibrary{calc}

\newtheorem{p-rule}{\bf Rule}

\usepackage{hyperref}
\usepackage{booktabs,tabularx}

\usepackage{booktabs,tabularx,array,ragged2e,pifont}
\newcolumntype{Y}{>{\RaggedRight\arraybackslash}X}
\newcolumntype{L}[1]{>{\RaggedRight\arraybackslash}p{#1}}
\newcommand{\cmark}{\ding{51}} 
\newcommand{\xmark}{\ding{55}} 

\usepackage{caption}      
\usepackage{dblfloatfix}  
\usepackage{placeins}     

\let\oldnl\nl
\newcommand{\nonl}{\renewcommand{\nl}{\let\nl\oldnl}}

\DeclarePairedDelimiter{\ceil}{\lceil}{\rceil}

\newcommand{\FuncName}[1]{\textsc{{#1}}}

\newcommand{\batch}{\mathcal{B}}

\newcommand{\sequence}{s}

\newcommand{\mem}{m}

\newcommand{\gen}{g}

\newcommand{\I}{\mathbb{I}}

\newif\ifFullVersion

\def\FullVersion{\let\ifFullVersion=\iftrue}
\def\ShortVersion{\let\ifFullVersion=\iffalse}
\FullVersion

\newcommand{\Azure}{\FuncName{{AzureConv}}\xspace}
\newcommand{\LongForm}{\FuncName{{LongForm}}\xspace}
\newcommand{\SelfConsistency}{\FuncName{{SelfConsistency}}\xspace}

\newcommand{\ReasonTwo}{\FuncName{{Reason2}}\xspace}

\newcommand{\ChatFift}{\FuncName{{Chat15}}\xspace}

\newcommand{\Ours}{\FuncName{{InferMax}}\xspace}

\newcommand{\Gurobi}{\FuncName{Gurobi}\xspace}
\newcommand{\Orca}{\FuncName{Orca}\xspace}
\newcommand{\vLLM}{\FuncName{vLLM}\xspace}
\newcommand{\vLLMS}{\FuncName{vLLM-Sys}\xspace}
\newcommand{\SARATHI}{\FuncName{Sarathi}\xspace}
\newcommand{\Sarathi}{\FuncName{Sarathi}\xspace}
\newcommand{\Splitwise}{\FuncName{Splitwise}\xspace}
\newcommand{\SGLang}{\FuncName{SGLang}\xspace}

\newcommand{\vLLMHY}{\FuncName{vLLM}$_{hy}$\xspace}
\newcommand{\SarathiPC}{\FuncName{Sarathi}$_{C=S}$\xspace}
\newcommand{\SarathiNOCP}{\FuncName{Sarathi}$_{nocp}$\xspace}

\newcommand{\InstInfer}{\FuncName{InstInfer}\xspace}
\newcommand{\LLMViewer}{\FuncName{LLM-Viewer}\xspace}
\newcommand{\DeepSpeed}{\FuncName{DeepSpeed}\xspace}
\newcommand{\FlexGen}{\FuncName{FlexGen}\xspace}

\newcommand{\Dynamo}{\FuncName{Dynamo}\xspace}
\newcommand{\DistServe}{\FuncName{DistServe}\xspace}
\newcommand{\Dejavu}{\FuncName{Dejavu}\xspace}
\newcommand{\ExeGPT}{\FuncName{ExeGPT}\xspace}
\newcommand{\AptServe}{\FuncName{AptServe}\xspace}
\newcommand{\vTensor}{\FuncName{vTensor}\xspace}

\newcommand{\InfiniGen}{\FuncName{InfiniGen}\xspace}
\newcommand{\NanoFlow}{\FuncName{NanoFlow}\xspace}
\newcommand{\Vidur}{\FuncName{Vidur}\xspace}

\newcommand{\MemServe}{\FuncName{MemServe}\xspace}
\newcommand{\FastServe}{\FuncName{FastServe}\xspace}
\newcommand{\ConServe}{\FuncName{ConServe}\xspace}
\newcommand{\CacheOPT}{\FuncName{CacheOPT}\xspace}
\newcommand{\DLPM}{\FuncName{D\textsuperscript{2}LPM}\xspace}
\newcommand{\KIVI}{\FuncName{KIVI}\xspace}
\newcommand{\KVzip}{\FuncName{KVzip}\xspace}
\newcommand{\HashEvict}{\FuncName{HashEvict}\xspace}
\newcommand{\Medha}{\FuncName{Medha}\xspace}
\newcommand{\Mooncake}{\FuncName{Mooncake}\xspace}
\newcommand{\BatchLLM}{\FuncName{BatchLLM}\xspace}
\newcommand{\FastSwitch}{\FuncName{FastSwitch}\xspace}
\newcommand{\LayerKV}{\FuncName{LayerKV}\xspace}
\newcommand{\SynergySched}{\FuncName{SynergySched}\xspace}

\newcommand{\KVPR}{\FuncName{KVPR}\xspace}
\newcommand{\LMCache}{\FuncName{LMCache}\xspace}
\newcommand{\KVShare}{\FuncName{KVShare}\xspace}
\newcommand{\NEO}{\FuncName{NEO}\xspace}
\newcommand{\HeadInfer}{\FuncName{HeadInfer}\xspace}

\newcommand{\FlashAttention}{\FuncName{FlashAttention}\xspace}

\SetKw{Continue}{continue}

\newcommand{\CASE}[1]{\STATE \textbf{case} #1\textbf{:} \begin{ALC@g}}
\newcommand{\ENDCASE}{\end{ALC@g}}

\newcommand{\DEFAULT}{\STATE \textbf{default:} \begin{ALC@g}}
\newcommand{\ENDDEFAULT}{\end{ALC@g}}
\newcommand{\DEFAULTLINE}[1]{\STATE \textbf{default:} }

\newcommand{\bluecomment}[1]{}

\newcommand{\kkim}[1]{#1}
\newcommand{\nexttodo}[1]{#1}
\newcommand{\eslliu}[1]{#1}

\newcommand{\todo}[1]{#1}
\newcommand{\xxx}[1]{#1}

\newcommand{\mlsys}[1]{#1}

\ifFullVersion

\else

\fi

\LinesNumbered
\SetAlgoCaptionSeparator{.}
\SetKwProg{Fn}{Function}{}{end}
\SetKwFor{uForEach}{foreach}{do}{}
\SetStartEndCondition{ (}{) }{)}
\SetNlSty{texttt}{}{:}
\SetArgSty{}

\usepackage{fancyhdr}
\newcommand{\shorttitle}{Review} 
\begin{document}

\title{Saving GPU Hours in LLM Inference System Development and Online Workloads with Simulation and DBMS-Inspired Cache Replacement Policies
}



\author{Kyoungmin Kim         \and
        Jiacheng Li \and
        Kijae Hong \and
        Qunyou Liu \and
        Darong Huang \and
        Anastasia Ailamaki
}


\institute{Kyoungmin Kim \at
              EPFL, Switzerland \\
              \email{kyoung-min.kim@epfl.ch}           
           \and
           Jiacheng Li \at
              NUS, Singapore \\
              \email{jiacheng.li@u.nus.edu}
           \and
           Kijae Hong \at
              POSTECH, South Korea \\
              \email{kjhong@dblab.postech.ac.kr}
           \and
           Qunyou Liu \at
              EPFL, Switzerland \\
              \email{qunyou.liu@epfl.ch} 
           \and
           Darong Huang \at
              EPFL, Switzerland \\
              \email{darong.huang@epfl.ch}
           \and
           Anastasia Ailamaki \at
              EPFL, Switzerland \\
              \email{anastasia.ailamaki@epfl.ch} 
}

\date{Received: November 2025 / Reviews out: April 2026}

\maketitle

\begin{abstract}
LLMs are increasingly used \todo{world-wide from daily tasks to agentic systems and data analytics,}
requiring significant GPU resources.
\kkim{While LLM inference systems are capable of serving millions of requests from multiple users, they often lack theoretical models to determine whether they achieve the performance upper bounds of underlying hardware resources. 
Beyond online workload serving, merely analyzing existing systems—or developing yet another one—is both GPU-intensive and labor-intensive.}
This paper provides a comprehensive survey of LLM inference systems, focusing on their cache management policies and availability.
We then show that simulations can be an effective tool to save GPU hours in the development and analysis phase of inference systems, revealing useful insights for developing better inference techniques, unlike how existing studies used simulations to find the best parameters inside a given system.
Finally, we provide theoretical tools to estimate the optimal performance and formulate new ideas.
Based on the theoretical analysis, especially on the cache management in LLM inference, we propose a simple yet effective cache replacement policy that can be easily plugged into existing preemptive schedulers and systems. We show that such a simple policy inspired from database systems can substantially save GPU hours in actual inference systems on online workloads.
{We share our experience submitting a journal paper to a database venue in November 2025 for anyone considering a similar path} (\href{https://docs.google.com/document/d/18A-wNwwJgXJtmM-KYhMVrwBqFazUa-TzouMZ785KvRo}{link}).


\end{abstract}

\section{Introduction}\label{sec:introduction}

Unlike LLM training, which is a one-time cost, LLM inference is an ongoing and significantly more expensive process due to high GPU costs and energy consumption. For instance, ChatGPT \kkim{approximately receives 4.6 billion user visits per month and 2.5 billion prompts per day}\footnote{https://explodingtopics.com/blog/chatgpt-users}, incurring GPU expenses that might exceed \$160 million monthly\footnote{https://seo.ai/blog/how-many-users-does-chatgpt-have, https://www.semianalysis.com/p/the-inference-cost-of-search-disruption}.
\kkim{Our back-of-the-envelope calculation also expects millions of dollars for monthly GPU expenses}\footnote{
We follow the assumptions in https://epoch.ai/gradient-updates/how-much-energy-does-chatgpt-use: GPT-4o has around 400B total parameters and 100B active parameters, the amount of compute per token is 2 FLOPs $\times$ active parameters, the average number of tokens is 500 per query, so each query uses $2 \times 10^{11} \times 500 \approx 10^{14}$ FLOPs. The effective FLOPs/s for H100-class GPUs is $9.9 \times 10^{13}$, which gives us $\frac{10^{14}}{9.9 \times 10^{13}} \approx 1$ second per query. Since H100 costs \$2-\$3/hour, 2.5 billion queries/day $\times$ 1/3600 hour $\times$ \$2-\$3/hour gives \$1.4-\$2.1 million per day, or \$42-\$63 million per month.
}.
Recent advancements, such as OpenAI's o1/o3 models, Deep Research\footnote{https://openai.com/index/introducing-deep-research}, and \kkim{agentic frameworks}, further underscore the rising demands of LLM inference, as these models extend inference time to enhance reasoning capabilities. Meanwhile, open-source models like Llama and DeepSeek enable broader adoption of LLM inference. Reducing LLM inference latency is thus crucial for both economic and environmental sustainability, given the substantial CO2 emissions associated with GPU usage \cite{Splitwise, CO2}.

\todo{LLMs are also increasingly making synergies with database systems and applications in either LLM4DB or DB4LLM fashion} \cite{LLMMeetDB}, \kkim{alongside the recent proliferation of AI4DB and DB4AI studies \cite{AIDBsurvey}.}
For \todo{LLM4DB}, LLMs are used in complex data management \cite{SIGMOD2024tutorial, LLM4DB}, database administration \cite{LLMasDBA, LLMDatabaseTuningManualDBBERT}, query optimization \cite{LLMQueryOptLLMR2, LLMQueryOptUnreasonable}, text-to-SQL translation \cite{multipathchasesql2024}, and \todo{semantic operators that extends relational operators to use LLMs \cite{LOTUS}.}


\todo{For DB4LLM,} researchers have focused on designing more efficient LLM inference systems utilizing database and operating systems techniques \cite{Orca, vLLM, DistServe, InfiniGen, NanoFlow, vTensor, InstInfer, KVCache, AptServe, SARATHI, SARATHI_SERVE, DeepSpeed}.
\kkim{The least recently used (LRU) cache replacement policy is widely adopted in many inference systems \cite{vLLM, SGLang, Mooncake, Dynamo} as for the buffer management in database systems \cite{LRU}.}

\kkim{DB4LLM and system development} efforts often involve redesigning and reconstructing entire \kkim{system stacks}. Such systems 
are labor- and hardware-intensive, requiring extensive analysis, implementation, and GPU usage.
\kkim{System development typically focuses on the final deliverable in terms of GPU hour savings in online workloads \cite{vLLM, SGLang, DeepSpeed}, but often ignores the GPU hours and costs spent in the development and analysis phase.}
\kkim{A recent study \cite{Vidur} shows that just finding the best set of system configuration knobs can take 42K GPU hours with an estimated cost of \$218K, while the GPU hours spent on development and analysis can cost much more.}\footnote{Unfortunately, we found no literature addressing the actual GPU hours and costs spent in development and analysis. Based on our experience on analyzing LLM inference times on real-world workloads, it required more than using 1K GPU hours with \$2K per week. The numbers from industry and actual system developments can be much higher, and also considering the massive number of research teams working independently.}
Moreover, it remains unclear whether existing systems are fully utilized, potentially leaving untapped optimization opportunities. A more cost-effective, goal-oriented approach is needed instead of relying on a naive trial-and-error process \kkim{consuming extensive GPU hours.}

\todo{As a first step toward a cost-effective approach, we adapt \kkim{simulation-based LLM inference} based on simple cost models, like the ones used in DMBSs, to predict inference performance without extensive use of GPUs, following the principles of \cite{Vidur}.} \kkim{However, unlike \cite{Vidur} that uses simulation to tune system configuration knobs, we use simulation to (a) \emph{analyze} LLM inference performance, including a what-if analysis that assumes unlimited hardware capacity to estimate performance upper bounds, and (b) \emph{develop} new strategies that can optimize the performance when \emph{deployed} to actual LLM inference systems.
Hence, we focus on saving GPU hours in both pre-deployment (development and analysis) and post-deployment phases.
We also show that simpler cost models, even linear ones, are effective enough to reveal interesting insights to derive such strategies, and their closed-form formula and monotonicity property enable principled and theoretical analysis in contrast to typical empirical evaluations.
}

\kkim{In this paper, we specifically target request preemption and cache replacement policies under high request contention. LLM inference systems cache each request's intermediate results in GPU memory whose size grows over time, as more output tokens are generated and their results are cached. Therefore, when the contention is high, the limited cache size may preempt some running requests and restart them later, freeing their cached data to prioritize running other requests.
While most studies focus on \emph{scheduling} requests (forming batches of requests and deciding when to launch each batch, analogous to multi-query optimization in DBMS), preemption and cache replacement policies remain largely underexplored. Existing systems typically rely on simple heuristics: using LRU on request completion to manage cached data, and preempting the newest request first (called NRF) when preemption occurs \cite{vLLM, SGLang}. Moreover, systems typically avoid preemption altogether by delaying request scheduling, striving to reduce contention and improve inference performance.
}

Contrary to popular belief, preemption should not be avoided. Our analysis based on theoretical and actual evaluations reveals a counterintuitive insight:  harnessing preemption can improve performance more than conservatively avoiding any preemption, particularly for short requests. 
We prove the optimality of preemption and non-preemption theoretically, depending on the request length, \kkim{including \emph{when} to preempt a request and \emph{which} requests to preempt in order to improve performance.}
First, we apply the five-minute rule \cite{FiveMinuteRule} in DBMSs to LLM inference and calculate the break-even intervals of keeping cache items in GPU memory before evicting them,
\kkim{to determine which requests or data are immediately preemptable. We conclude that this rule justifies the non-preemption of any requests, but is insufficient to tell which requests should be preempted to improve performance.}
Instead, we define a concept of \emph{progress} and propose a \kkim{progress-aware} preemption and cache replacement policy, \kkim{which we call \emph{shortest-request first (SRF)} policy.} 
The \kkim{progress reflects the cost of processing a batch of request (lower the better) and the number of tokens generated from the batch (larger the better), considering the restarting overhead of requests. We show that this progress is large for preempting short requests.}

\kkim{Regarding the workloads,} we first target general LLM inference workloads with high demands such as conversations and text generations in this paper for maximum and universal GPU hour savings, \kkim{not database-specific workloads in LLM4DB applications.}
When deployed to \nexttodo{a real-world LLM inference system}, our policy consistently reduces inference latencies by \nexttodo{up to 15\%} without performance regression or losing fairness in serving concurrent requests, where the estimated optimal reduction is \nexttodo{40\%} with an infinite cache size. 
From a simple calculation, this indicates the potential for significant economic and environmental savings, for example, more than \nexttodo{\$24 million} per month for ChatGPT.
We also observe that LLM inference has hit a memory wall as in analytical query processing in DBMSs \cite{MemoryCentricDB}, where the bandwidth matters. 
This calls for a better implementation of hardware-conscious LLMs and algorithms inside LLMs to resolve the memory-boundness of LLM inference and hide data transfer overheads.




\todo{In summary, our contributions are as follows.}
\begin{itemize}[topsep=0pt, leftmargin=0.5cm]
  \item \kkim{We provide a comprehensive survey of LLM inference systems from the perspective of data management.}
  \item \kkim{We show that simulations based on simple cost models are sufficient to yield meaningful insights that lead to the development of new inference performance optimization techniques without requiring many GPU hours during the development and analysis phase.}
  \item We find that preempting short requests under high memory contention can improve performance by \xxx{2x} when compared to the common practice of avoiding preemption altogether. The reason is that short requests incur low restart overheads and hence generate output tokens faster despite being preempted. We prove our findings through theoretical formulations.
  \item We apply Gray and Putzolu's five-minute rule to cache management in LLM inference, and show that the break-even interval for keeping cache items in GPU memory varies from sub-second to a few minutes, depending on the request length, \kkim{exceeding the actual reuse intervals.}
  \item We propose a new preemption and cache replacement policy that preempts the shortest requests first. It improves inference performance by \nexttodo{up to 15\%} under high contention compared to preempting newest requests first in \nexttodo{a real-world inference system, \vLLM \cite{vLLM}, and also improves performance in other systems, \SGLang \cite{SGLang}, and \Dynamo \cite{Dynamo}, by up to 20\%.} 
  Our policy is easy to implement with a few lines of code in \kkim{actual systems} and shows \xxx{no performance regression or losing fairness}. These indicate \kkim{a safe approach to} substantial GPU cost savings without extensive \kkim{development} overheads.
\end{itemize}

\kkim{The rest of the paper includes background through a survey of LLM inference systems focusing on data management (Section \ref{sec:background}), an overview of our simulation-based analysis and actual system integration framework (Section \ref{sec:overview}), 
the simulation design (Section \ref{sec:sim_design}) and results (Section \ref{sec:sim_result}), a five-minute rule for LLM inference (Section \ref{sec:five_minute_rule}), a theoretical analysis and our new cache replacement policy (Section \ref{sec:theory}), the evaluation of our policy in actual LLM inference systems (Section \ref{sec:deploy}), and a conclusion (Section \ref{sec:conclusion}).
}

\section{Background}\label{sec:background}





\kkim{This section provides background information about processing LLM inference requests and cache management, analogy between LLM inference systems and DBMSs, and a comprehensive survey of related work.}

\subsection{LLM Inference Request Lifetime and Cache Management}\label{subsec:background:request_lifetime}

\begin{figure}[h!]
\centering
\includegraphics[width=0.92\columnwidth]{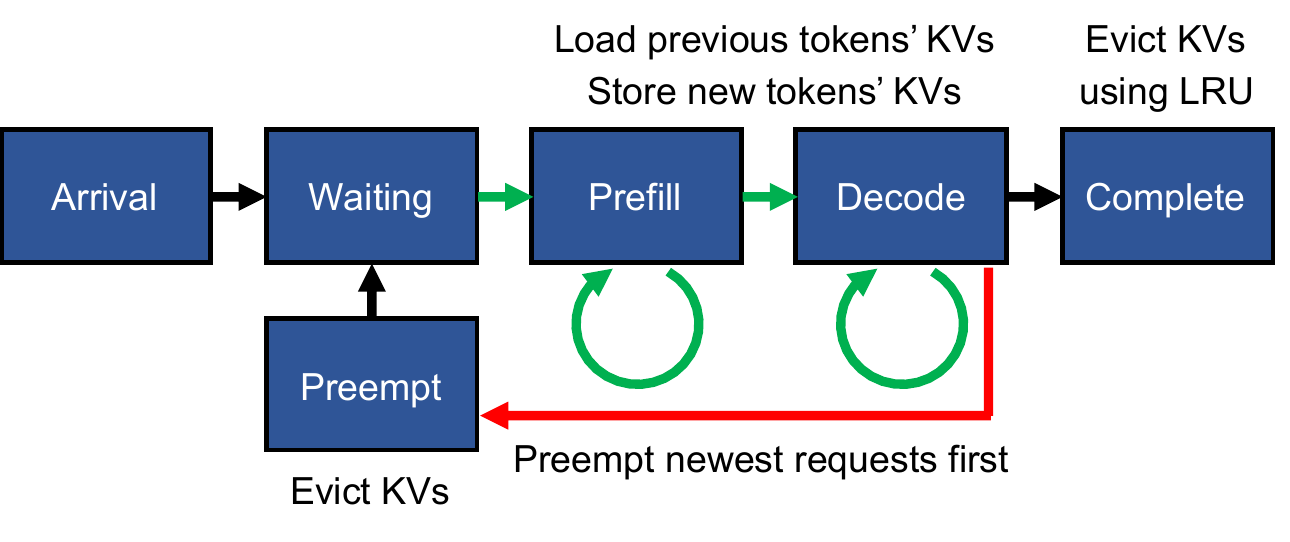}
\caption{\kkim{Life cycle of an LLM inference request and management of its intermediate results (key-value tensors or KVs). Our main focus is on the decode-to-preempt transition (red arrow) and its cache replacement policies, which are orthogonal to and can be plugged into preemptive schedulers in LLM inference systems that largely focus on the waiting-prefill-decode transitions (green arrows).}
}\label{fig:lifetime}
\end{figure}

\kkim{As shown in Figure \ref{fig:lifetime}, when an LLM inference request arrives to an inference system, it is first kept in a waiting queue until the request \emph{scheduler} launches a batch containing this request. In the \emph{prefill} phase, the request's input tokens or prompt tokens are processed by an LLM. The prompt can be chunked and processed in multiple steps/batches. If the prefill is done, it generates a new output token and transits to the decode phase. In each \emph{decode} step, the last generated token is processed to generate another one. When the available resources fall short, some running requests can be \emph{preempted} to prioritize running others. These are restarted or \emph{refilled} later. When the last generated token is EOS (end-of-sequence) token or the output length reaches a specified limit, the request is completed.}

\kkim{In the mainstream LLM models, which are \emph{autoregressive} Transformer-based models we consider in this paper, each attention operation inside a model's layer produces one \emph{key-value (KV)} tensor for each processed token \cite{attention}. These KVs are maintained in the \emph{KV cache} \cite{KVCache} in GPU memory and reused for all subsequent decode steps, since 1) processing later tokens requires the KVs of all preceding tokens, and 2) it is more efficient to cache and reuse KVs than recomputing them every step. The attention can be expressed mathematically as}


\begin{equation}\label{eq:attention_formula}
\begin{split}
    & Q_{i} = f_Q(x_{i}), K_{1:i} = f_K(x_{1:i}), V_{1:i} = f_V(x_{1:i}), \\ 
    & z_{i} = Attention(Q_{i}, K_{1:i}, V_{1:i}),
\end{split}
\end{equation}

\noindent \kkim{where $x_i$ is the hidden state of the $i$-th token of a request, $Q_{i}$ is the query tensor for the token, and $K_{1:i}$/$V_{1:i}$ is the key/value tensors for the first to $i$-th tokens. Each $f$ denotes a matrix multiplication. $Attention$ is the attention operation generating another hidden state, $z_i$, for downstream intra-layer operations. Note that $K_{1:i}$ and $V_{1:i}$ can be reused for processing any subsequent $j$-th token $x_j (j > i)$. $z_i$ is transformed into the input hidden state $x_i$ of the next layer, making the computation of $x_i$ autoregressive (depending on all preceding tokens). We detail the model layers in Section \ref{subsec:sim_design:cost_model}.}

\kkim{Since every running request appends one KV to the cache in GPU memory at each decode step, where the GPU memory has limited capacity,
some requests may be preempted and their KVs freed if the cache becomes full.
When a request is completed, its KVs can be kept for future requests with the same prefix and  replaced under the LRU policy \cite{vLLM, SGLang}.}


\kkim{Refilling the KVs can be done either by 1) recomputing KVs in GPU or 2) offloading KVs to secondary storage (e.g., CPU memory, SSD) and swapping in when they are needed.
While the offloading requires fast interconnects between hardware devices, low PCIe bandwidths compared to the high computational capacity of GPUs often increase the end-to-end latencies of inference requests. A popular inference system, \vLLM \cite{vLLM}, deprecates offloading because of higher latencies and complex data management. 
Offloading can be a reasonable option \cite{Dynamo, Mooncake} with faster interconnects, newer hardware chips (e.g., Grace-Hopper superchips), or to avoid refilling long inputs where prefill/refill takes quadratic time to request lengths (Section \ref{subsec:sim_design:cost_model}), reduce GPU usage \cite{Dynamo}, and reuse KVs across requests and GPUs \cite{Mooncake}. 
The overhead of transferring KVs may dominate runtime if not overlapped carefully \cite{ren2025characterizing}.
}
\kkim{We develop our ideas in Section \ref{subsec:theory:which_request_preempt} based on that both recomputation and swapping-in costs monotonically increase with the number of KVs (Sections \ref{subsec:sim_design:cost_model} and \ref{subsec:sim_result:batch_swap_recompute}).}

\kkim{Other than the attention operation and KVs, the matrix multiplications in each LLM model layer require loading model weights (matrices) from GPU memory. To amortize this overhead, the request scheduler (simply scheduler) batches multiple requests together.}

\subsection{Analogy between DBMSs and LLM Inference Systems}\label{subsec:background:analogy}

\kkim{Before we survey related work, we set an analogy between LLM inference systems and DBMSs for a better understanding of inference systems and clarification of our scope. Here, we consider \emph{online} inference systems that schedule concurrent requests from multiple users, not pre-planned \emph{offline} inference systems that are specialized in certain optimizations, such as supporting LLMs larger than the GPU memory \cite{FlexGen}, partial KV offloading to quickly approximate the attention operation \cite{InfiniGen}, and KV quantization \cite{KIVI}.
}

\begin{figure*}[h!]
\centering
\includegraphics[width=1.7\columnwidth]{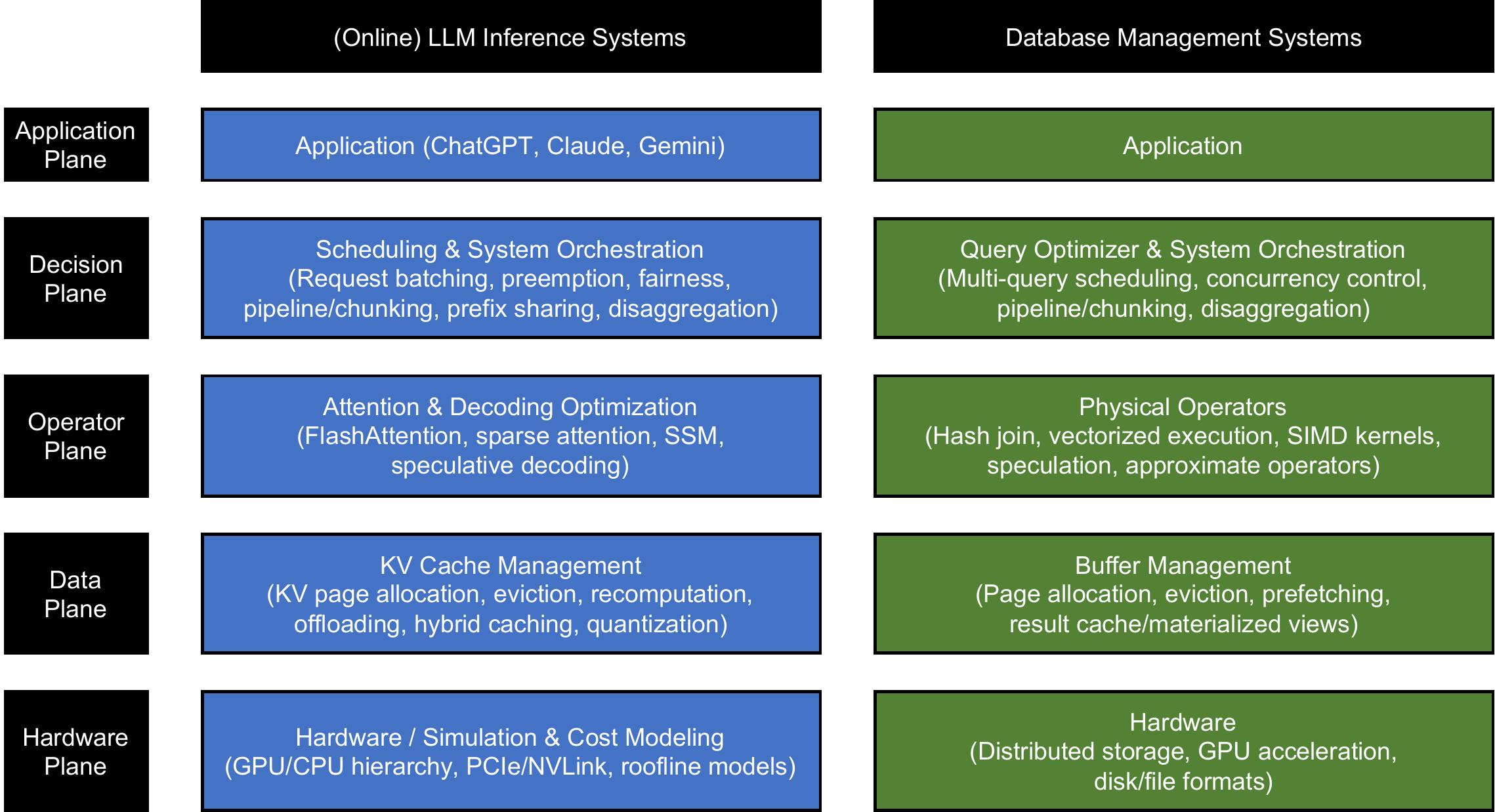}
\caption{Analogy between LLM inference systems and DBMSs in five layers/planes, each box listing examples.}\label{fig:layers}
\end{figure*}

\kkim{Figure \ref{fig:layers} shows five planes -- application, decision, operator, data, and hardware -- in both types of systems.
At the highest level, applications refer to the services accessible to users. The decision plane corresponds to request scheduling for processing concurrent requests while maximizing system performance, considering the available hardware and software resources. This is similar to the query optimization in DBMSs.
For the operator plane, the attention operation and optimized decoding strategies are analogous to physical operator implementations in DBMSs.
For the data plane, KV cache management is akin to the buffer management, LRU being the dominant strategy to maintain the completed requests' KVs.
Hardware includes utilizing available resources. 
}

\kkim{This paper focuses on request preemption (decision plane) and KV cache replacement policies (data plane), the decode-to-preempt transition in Figure \ref{fig:lifetime}.
This scope is largely orthogonal to other extensively-studied scheduling techniques (decision plane) that consider the waiting-prefill-decode transitions in Figure \ref{fig:lifetime}, but can be plugged in to any preemptive schedulers.
The following section explains each plane in detail.
}

\subsection{Related Work}\label{subsec:background:related}







\kkim{This section surveys LLM inference systems through the lens of the five planes in Figure \ref{fig:layers} and summarizes them in the taxonomy in Table \ref{tab:llm-taxonomy}.} From the database domain, our what-if analysis \kkim{in Section \ref{sec:sim_result}} resembles the one in \cite{WhatIfDB}, but ours focuses on LLM inference.
We focus on the closest domains since LLM inference system and optimization is a very broad field \cite{LLMMeetDB, LLMInferenceSurvey}. \kkim{We also focus on systems with open-sourced implementations whenever possible to facilitate transparent reproducibility and robust follow-up research.}

\kkim{Overall, we observe that the first wave of general-purpose, open-source inference engines such as \vLLM \cite{vLLM}, \SGLang \cite{SGLang}, \SARATHI \cite{SARATHI}, and \DistServe \cite{DistServe} has largely defined the \emph{baseline} system design space. Subsequent systems tend to introduce incremental scheduling or KV-management techniques on top of these baselines, but there is still no widely accepted, ground-breaking redesign of the serving stack. In particular, preemption and KV-cache replacement under contention are much less explored than batching and operator-level optimizations. Our work targets exactly these two aspects (decision and data planes) and shows that even simple, theoretically grounded policies can still yield substantial GPU-hour savings when integrated into these mature stacks.}

\subsubsection{Application Plane}

\kkim{At the application plane, LLM inference systems serve diverse workloads such as conversational assistants, code assistants, RAG pipelines, and data-centric applications such as semantic query processing and agents over databases \cite{LOTUS}. These applications impose heterogeneous service-level objectives (SLOs): interactive chat requires low time-to-first-token (TTFT) and low time-per-output-token (TPOT); batch analytics workloads emphasize throughput and GPU-hour cost; and multi-tenant deployments must ensure fairness across users and models.
LLM4DB applications mentioned in Section \ref{sec:introduction} fall into this plane. Our work is mostly orthogonal to application-level design: we do not change the user-visible interface or the semantics of the LLM outputs.
}

\subsubsection{Decision Plane: Scheduling and System Orchestration}


\kkim{At the decision plane, the main goal is to schedule concurrent requests and orchestrate model execution to maximize throughput and SLO attainment under hardware constraints. Most existing systems focus on batching and phase scheduling.}

\noindent \underline{\textbf{Continuous and hybrid batching.}} 
\kkim{\Orca \cite{Orca} introduces continuous batching, which breakdowns a request lifetime into multiple steps in Figure \ref{fig:lifetime} and schedules requests step-wise. \SARATHI \cite{SARATHI, SARATHI_SERVE} and \DeepSpeed \cite{DeepSpeed} propose chunked prefill and hybrid prefill–decode batching: large input prompts are processed in chunks, and prefill and decode tokens are interleaved in the same batch to better utilize GPU resources.}

\noindent \underline{\textbf{Prefill–decode disaggregation.}} 
\kkim{\DistServe \cite{DistServe}, \Dejavu \cite{Dejavu}, \Splitwise \cite{Splitwise}, \ExeGPT \cite{ExeGPT}, and \Dynamo \cite{Dynamo} disaggregate the prefill and decode phases across GPUs. Prefill GPUs handle heavy one-shot context processing, while decode GPUs specialize in long auto-regressive decodes. 
These systems must decide how to route KVs between prefill and decode workers.}


\noindent \underline{\textbf{Pipelining and intra-batch parallelism.}} 
\kkim{\NanoFlow \cite{NanoFlow} further partitions each batch into \emph{nano-batches} to overlap operators within a layer and across requests, while \Medha \cite{Medha} adopts sequence pipeline parallelism and adaptive prefill chunking.}




\noindent \underline{\textbf{Fairness and predictive scheduling.}}
\kkim{\FastServe \cite{FastServe} proposes multi-level-queue scheduling with proactive KV offloading for latency-critical requests. \ConServe \cite{ConServe} designs SLO-aware token-level scheduling with layer-wise preemption. \DLPM \cite{D2LPM} focuses on locality-aware fair scheduling, while \SynergySched \cite{SynergySched} combines predictive models with a two-layer scheduler. These works emphasize fairness and SLO satisfaction but do not deeply study the interaction between preemption and KV-cache replacement.}

\noindent \underline{\textbf{Output-length-aware scheduling.}}
\kkim{Several studies predict or rank output lengths to improve scheduling \cite{OPrediction, OPrediction2, ORank}, which is out of our scope. Our approach complements such work: through simulation and hypothetical schedulers that leverage exact output lengths, we quantify the performance upper bound of such predictors can provide, similar to evaluating the impact of perfect cardinality estimates on query plans before designing new estimators \cite{JOB}.}
Furthermore, our findings also apply to scenarios where requests have identical lengths.

\kkim{Across these systems, the preemption policy under KV-cache contention is typically very simple: most open-source implementations adopt newest-request first (NRF) for preemption and keep completed requests' KVs under the LRU policy (Table \ref{tab:llm-taxonomy}). To the best of our knowledge, our work, starting from our technical report \cite{kim2024faster}, is the first to propose the shortest-request-first (SRF) preemption and cache replacement policy.}

\FloatBarrier

\begin{table*}[!htbp]
\centering
\small
\renewcommand{\arraystretch}{1.15}
\setlength{\tabcolsep}{4pt}
\caption{\kkim{Taxonomy of LLM inference systems and simulators. ``KV Repl. (P)'' and ``KV Repl. (C)'' denote the KV cache replacement policy under \textbf{P}reemption and \textbf{C}ompletion (Figure \ref{fig:lifetime}) in open-source implementations with options: NRF (newest request first), 
SRF (shortest request first), LRU (least recently used), and `*' (wildcard). 
``KV Refill'' indicates whether KVs are recomputed or offloaded then swapped in after being evicted/replaced. 
KVs can also be offloaded simply after request completions for sharing across requests or hardware devices.
}}
\label{tab:llm-taxonomy}
\begin{tabularx}{\textwidth}{
  L{3.0cm} L{2.1cm} Y L{0.9cm} L{0.9cm} L{1.6cm} L{1.1cm}}
\toprule
\textbf{System / Paper} & \textbf{Main Plane} &
\textbf{Key Techniques} &
\textbf{KV Repl. (P)} & \textbf{KV Repl. (C)} & \textbf{KV Refill} & \textbf{Open-sourced} \\
\midrule

\textbf{\Orca} \cite{Orca} & Scheduling &
Continuous batching &
-- & -- & -- & \xmark \\

\textbf{\vLLM} \cite{vLLM} & Attention, Cache Mgmt. &
Paged attention, paged KV cache &
NRF & LRU & Recomp., offload & \cmark \\

\textbf{\SGLang} \cite{SGLang} & Attention, Cache Mgmt. &
Radix attention for shared prefixes &
NRF & LRU & Recomp., offload & \cmark \\


\textbf{\SARATHI} \cite{SARATHI, SARATHI_SERVE} & Scheduling &
Chunked prefill, hybrid prefill-decode batching &
NRF & \xmark & Recomp. & \cmark \\


\textbf{\DistServe} \cite{DistServe} & Scheduling &
Prefill–decode disaggregation &
NRF & \xmark & Offload & \cmark \\

\textbf{\Splitwise} \cite{Splitwise} & Scheduling &
Prefill–decode disaggregation &
-- & -- & -- & \xmark \\

\textbf{\ExeGPT} \cite{ExeGPT} & Scheduling &
Prefill–decode disaggregation, constraint-aware scheduling &
-- & -- & -- & \xmark \\

\textbf{\Mooncake} \cite{Mooncake} & Scheduling, Cache Mgmt. &
Prefill-decode disaggregation, disaggregated KV cache pool, prediction-based early rejection &
\xmark & LRU & Offload & \cmark \\

\textbf{\Dynamo} \cite{Dynamo} & Scheduling &
Prefill–decode disaggregation, distributed scheduling &
NRF & LRU & Recomp., offload & \cmark \\

\textbf{\NEO} \cite{NEO} & Scheduling &
Asymmetric GPU-CPU pipelining &
NRF & \xmark & Offload & \cmark \\


\textbf{\NanoFlow} \cite{NanoFlow} & Scheduling, Cache Mgmt. &
Nano-batching for intra-batch pipelining, asynchronous scheduling &
\xmark & \xmark & Offload & \cmark \\


\textbf{\AptServe} \cite{AptServe} & Cache Mgmt. &
Hybrid KV/activation cache &
NRF & LRU & Recomp. & \cmark \\


\textbf{\vTensor} \cite{vTensor} & Attention, Cache Mgmt. &
Decoupled KV allocation and attention operation &
-- & -- & -- & \xmark \\


\textbf{\InstInfer} \cite{InstInfer} & Attention, Cache Mgmt. &
KV offload to flash storage &
-- & -- & -- & \xmark \\


\textbf{\MemServe} \cite{MemServe} & Scheduling Cache Mgmt. &
Memory pool for distributed KV cache, locality-aware scheduling &
-- & -- & -- & \xmark \\

\textbf{\FastServe} \cite{FastServe} & Scheduling &
Multi-level-queue scheduling, proactive offloading &
-- & -- & -- & \xmark \\

\textbf{\ConServe} \cite{ConServe} & Scheduling &
SLO-aware token-level scheduling, layer-wise preemption &
-- & -- & -- & \xmark \\

\textbf{\CacheOPT} \cite{CacheOPT} & Scheduling, Cache Mgmt. &
Output-length prediction for proactive KV allocation, runtime KV recompute/offload selection &
-- & -- & -- & \xmark \\

\textbf{\DLPM} \cite{D2LPM} & Scheduling &
Locality-aware fair scheduling &
-- & -- & -- & \xmark \\

\textbf{\HashEvict} \cite{HashEvict} & Attention, Cache Mgmt. &
LSH-based eviction for low-attention tokens &
-- & -- & -- & \xmark \\

\textbf{\Medha} \cite{Medha} & Scheduling &
Adaptive prefill chunking, sequence pipeline parallelism &
-- & -- & -- & \xmark \\

\textbf{\BatchLLM} \cite{BatchLLM} & Scheduling &
Global prefix sharing, decode prioritization &
-- & -- & -- & \xmark \\

\textbf{\FastSwitch} \cite{FastSwitch} & Scheduling, Cache Mgmt. &
KV page grouping &
-- & -- & -- & \xmark \\

\textbf{\LayerKV} \cite{LayerKV} & Scheduling, Cache Mgmt. &
Layer-wise KV allocation &
-- & -- & -- & \xmark \\

\textbf{\SynergySched} \cite{SynergySched} & Scheduling &
Predictive scheduling, two-layer scheduling &
-- & -- & -- & \xmark \\

\textbf{\LMCache} \cite{LMCache} & Cache Mgmt. &
Distributed KV cache layer &
\xmark & \xmark & Offload & \cmark \\

\textbf{\KVShare} \cite{KVShare} & Scheduling, Cache Mgmt. &
Multi-tenant KV sharing &
-- & -- & -- & \xmark \\

\textbf{\DeepSpeed} \cite{DeepSpeed} & Offline System &
Chunked prefill, hybrid prefill-decode batching &
\xmark & \xmark & \xmark & \cmark \\

\bottomrule
\end{tabularx}
\end{table*}

\begin{table*}[!htbp]
\ContinuedFloat
\centering
\small
\renewcommand{\arraystretch}{1.15}
\setlength{\tabcolsep}{4pt}
\caption{\kkim{Taxonomy of LLM inference systems and simulators (continued).}}
\begin{tabularx}{\textwidth}{
  L{3.0cm} L{2.1cm} Y L{0.9cm} L{0.9cm} L{1.6cm} L{1.1cm}}
\toprule
\textbf{System / Paper} & \textbf{Main Plane} &
\textbf{Key Techniques} &
\textbf{KV Repl. (P)} & \textbf{KV Repl. (C)} & \textbf{KV Refill} & \textbf{Open-sourced} \\
\midrule


\textbf{\Dejavu} \cite{Dejavu} & Offline System &
Prefill–decode disaggregation, layer-wise KV streaming &
\xmark & \xmark & \xmark & \cmark \\

\textbf{\FlexGen} \cite{FlexGen} & Offline System &
Offline inference, larger model support than GPU memory &
\xmark & \xmark & \xmark & \cmark \\

\textbf{\InfiniGen} \cite{InfiniGen} & Offline System &
Partial layer-wise KV streaming &
\xmark & \xmark & \xmark & \cmark \\

\textbf{\KIVI} \cite{KIVI} & Offline System &
2-bit quantization of KV cache &
\xmark & \xmark & \xmark & \cmark \\

\textbf{\KVzip} \cite{KVzip} & Offline System &
KV importance selection, query-agnostic KV compression &
\xmark & \xmark & \xmark & \cmark \\

\textbf{\KVPR} \cite{KVPR} & Offline System &
Hybrid KV recomputation and offloading &
\xmark & \xmark & Offload & \cmark \\

\textbf{\HeadInfer} \cite{HeadInfer} & Offline System &
Head-wise KV offloading &
\xmark & \xmark & Offload & \cmark \\


\textbf{\Vidur} \cite{Vidur} & Scheduling, Simulation &
Learned cost models based on actual GPU times &
NRF & \xmark & \xmark & \cmark \\

\textbf{\LLMViewer} \cite{LLMViewer} & Simulation &
Theoretical cost models based on roofline models &
\xmark & \xmark & \xmark & \cmark \\

\textbf{Our Work} & Scheduling, Cache Mgmt., Simulation &
Learned cost models based on roofline models and actual GPU times, SRF integrated in online inference systems &
\kkim{SRF} & \kkim{*} & \kkim{*} & \cmark (planned) \\

\bottomrule
\end{tabularx}
\end{table*}

\subsubsection{Operator Plane: Attention and Decoding Optimization}

\kkim{At the operator plane, LLM inference systems optimize attention and decoding kernels, which are analogous to physical operator implementations in DBMSs.}

\kkim{A large body of work focuses on more efficient attention kernels. \FlashAttention \cite{FlashAttention, FlashAttention_2, FlashAttention_3} avoids unnecessary data movements during the attention computation without affecting attention outputs, which is now widely adopted and are enabled by default in systems such as \vLLM. Sparse attention variants \cite{SlidingWindowAttention, Linformer, team2025kimi} and state-space models (SSMs) \cite{SSM} change the attention structure and outputs to reduce complexity, but may degrade LLM accuracy \cite{SparseTradeOff, MambaStudy, OversmoothingSSM}.
We restrict ourselves to full-attention transformer models with \FlashAttention-style kernels, since 1) a survey \cite{AIReport} shows that transformer alternatives including SSM and hybrid models remain niche, and 2) sparse attention and SSMs simplify the KV management where requests may keep a constant number of KVs during the lifetime.
}


\kkim{Speculative decoding techniques \cite{SpeculativeDecoding} maintain a smaller proxy model that proposes candidate tokens, which the large model then verifies. This is akin to a work in DBMSs \cite{SpeculationDBMS}. The efficiency gains depend on the proxy model's accuracy and the composition of accepted versus rejected tokens. While speculative decoding can substantially improve efficiency, it complicates cost modeling and cache-management analysis, since speculative tokens may be discarded. We focus on single-model decoding and leave the integration of speculative decoding into our simulation and theoretical analysis as future work.}



\subsubsection{Data Plane: KV Cache Management}

\kkim{The data plane in LLM inference systems is centered around the KV cache: how KVs are allocated, laid out, evicted, recomputed, offloaded, and reused across requests and devices. This plane is directly analogous to buffer and result-cache management in DBMSs, where LRU and its variants dominate.}

\noindent \underline{\textbf{Paged and prefix-aware KV caches.}}
\kkim{\vLLM \cite{vLLM} introduces paged attention and a paged KV cache, which manage KVs with fixed-size pages, improving serving throughput. 
\vTensor \cite{vTensor} decouples the KV cache allocation and attention computation in \vLLM.
\SGLang \cite{SGLang} proposes radix attention, which effectively reuses the KVs and attention computations over the shared prefixes among multiple requests. These systems use LRU for completed requests and a simple NRF heuristic for preemption (Table \ref{tab:llm-taxonomy}).}


\noindent \underline{\textbf{Offloading during request lifetime vs. completion.}}
\kkim{It is useful to distinguish two different roles of KV offloading.
(i) Offloading during a request's lifetime is used to temporarily free GPU memory under contention and later swap KVs back to continue the same request. Due to the low PCIe bandwidth between CPU and GPU that makes KV offloading a critical bottleneck, \InfiniGen \cite{InfiniGen} offloads KVs to CPU memory and streams back only the most \emph{important} tokens to approximate the full attention result. 
\InstInfer \cite{InstInfer} offloads KVs to flash storage, and \NanoFlow \cite{NanoFlow} uses offloading as part of its intra-batch pipelining strategy. These techniques trade PCIe or NVMe bandwidth against GPU memory capacity and are highly sensitive to interconnect characteristics.
(ii) Offloading after request completion aims at KV reuse across future requests or across GPUs. \vLLM and \SGLang can store completed prefixes for multi-turn conversations or for global prefix sharing. \LMCache \cite{LMCache} builds a distributed KV cache layer; \KVShare \cite{KVShare} shares KVs across tenants; and \Mooncake \cite{Mooncake} maintains a disaggregated KV pool with prediction-based early rejection of low-value prefixes. Our work targets case (i): we focus on what to preempt and evict during a request's lifetime, while remaining compatible with post-completion KV reuse mechanisms.}

\noindent \underline{\textbf{Hybrid, layered, and token-level policies.}}
\AptServe \cite{AptServe} proposes a hybrid cache of KVs and the input activations for computing KVs to reduce memory footprint, but it can rather increase the footprint under grouped-query attention where KVs are smaller than activations \cite{GroupedQueryAttention}. 
\kkim{\FastSwitch \cite{FastSwitch} groups KV pages to improve spatial locality; \LayerKV \cite{LayerKV} allocates KVs layer-wise to better match memory hierarchies; \MemServe \cite{MemServe} and \LMCache \cite{LMCache} manage distributed KV pools across GPUs; and \KVPR \cite{KVPR} and \HeadInfer \cite{HeadInfer} explore hybrid recomputation-offloading and head-wise KV placement. \HashEvict \cite{HashEvict} uses locality-sensitive hashing to evict tokens that receive low attention, effectively moving from request-level to token-level eviction.}

\noindent \underline{\textbf{Compression and quantization.}}
\kkim{To further reduce memory footprint, \KIVI \cite{KIVI} applies 2-bit quantization to KVs, and \KVzip \cite{KVzip} performs query-agnostic KV compression based on importance.
These techniques are orthogonal to scheduling and cache replacement decisions and can be combined with our policies.}

\kkim{Despite this rich space of mechanisms, replacement policies under contention remain simple. As summarized in Table \ref{tab:llm-taxonomy}, open-source systems with online schedulers typically use (i) NRF for preemption 
and (ii) LRU for completed requests. 
To our knowledge, no prior work systematically studies the interaction between request length and preemption, nor proposes our SRF policy in \cite{kim2024faster}. Our SRF fills this gap: it chooses victims based on a progress metric that jointly captures KV size, restart overhead, and contribution to throughput, and it can be plugged into any preemptive scheduler and KV representation without changing other planes.}


\subsubsection{Hardware Plane: Hardware, Simulation, and Cost Modeling}


\kkim{The hardware plane concerns the underlying compute and memory hierarchy (GPUs, CPUs, interconnects, and storage) and how they are modeled. \LLMViewer \cite{LLMViewer} derives theoretical operator-level cost models directly from hardware roofline parameters (FLOPs/s and bandwidth) and model structure, and shows that such models generalize well across GPUs. However, it analyzes only single batches without scheduling, does not model chunked prefill or preemption, and does not connect its estimates back to end-to-end workload behavior.}


\kkim{\Vidur \cite{Vidur} is the first to provide a full-fledged LLM inference simulator with learned cost models based on measured GPU times. It allows service providers to explore hardware and system configurations under budget constraints, but its sklearn-based models (e.g., random forests) are not theoretically grounded, and their non-monotonicity makes it difficult to embed them into formal optimization frameworks such as our constraint satisfaction problem (CSP) \cite{CSP} in Section \ref{subsec:theory:csp}. \ExeGPT \cite{ExeGPT} also uses simulation to decide how to assign GPUs to prefills and decodes in a disaggregated setup. However, these focus primarily on configuration search rather than deriving new scheduling or cache replacement policies.}

\kkim{Our work connects these lines of research. We start from roofline-style theoretical models as in \LLMViewer, derive simple linear cost models in terms of token counts and KV sizes (Section \ref{subsec:sim_design:cost_model}), and then fit their coefficients using a few hours of profiling data as in \Vidur. We show that these linear models are accurate enough (6\% average error in Section \ref{sec:sim_result}) to capture end-to-end performance, generalize across GPU types, and, importantly, are closed-form and monotone. These properties allow us to translate scheduling into CSP 
to predict performance upper bounds and conduct theoretical analysis to derive new policies that can actually save GPU hours in online inference.
Not only the schedulers in existing systems, we also use hypothetical ones to predict performance upper bounds as mentioned earlier.
}

\section{Overview}\label{sec:overview}









This section introduces \kkim{an overview of our approach to reduce GPU hours spent in both pre-deployment (development and analysis) and post-deployment (inference request serving) phases in LLM inference systems.}

\begin{figure}[h!]
\centering
\includegraphics[width=1.0\columnwidth]{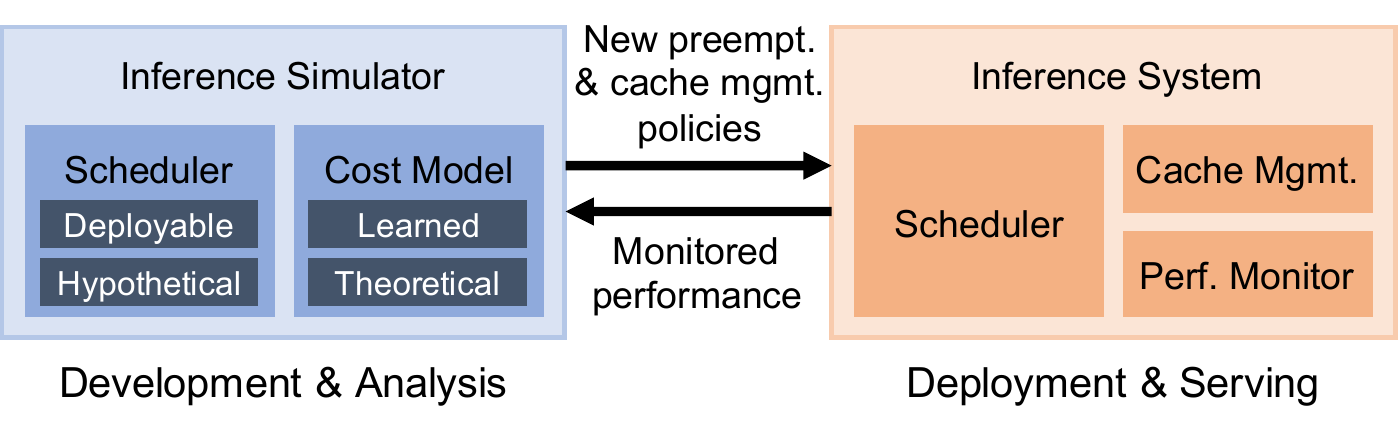}
\caption{\kkim{Bidirectional approach between inference simulators and systems to reduce GPU hours in both development and online inference serving phases.}}\label{fig:overview}
\end{figure}

\kkim{Figure \ref{fig:overview} illustrates this. The inference simulator captures and abstracts schedulers in real-world online inference systems, launching a batch of inference requests at each step. Each batch progresses the involving requests' states by one arrow in Figure \ref{fig:lifetime}.
Instead of using actual GPUs to process the batches, the simulator increments the global timer by the batch execution times estimated by the cost models. The estimation can be conducted in CPUs without using GPUs.
As each workload run can take hours of GPUs \cite{AzurePublicDataset}, this approach can save significant amount of GPU hours.
}

\kkim{The cost models are learned from performance profiling results \cite{Vidur} or theoretically built from hardware characteristics \cite{LLMViewer}.
We take a hybrid approach, starting from linear theoretical models and learn the coefficients from profiling results without altering the model structure.
Both learning and estimation occur in CPUs.
}

\kkim{While existing simulators focus on \emph{deployable} schedulers, i.e., implemented in inference systems, we also propose to use \emph{hypothetical} schedulers that allow estimating performance upper bounds leveraging the workloads information which is unknown in advance in online serving scenarios. For example, we can leverage the output lengths of requests to better schedule them considering their exact resource demands, preventing unnecessary resource contentions, and the resulting scheduler acts as a performance upper bound of the output-length-prediction schedulers in Section \ref{subsec:background:related}.
We can also conduct diverse what-if analyses, for example, decreasing or increasing the GPU memory size or bandwidth considering multi-tenancy and next-generation GPUs, which also act as performance lower and upper bounds.
}




\kkim{A benefit of using simulation is that the development overhead is much smaller inside simulators than in actual inference systems. For instance, the scheduler implementation of \vLLM \cite{vLLM} is around 2.1K lines of python code, which can be abstracted into 0.2K lines in simulators \cite{Vidur}.}


\kkim{By using simulations, we first show that we can save GPU hours in the development and analysis phase, where simple linear cost models are enough to capture actual inference performance and provide meaningful insights. They also provide a closed-form of cost estimation and monotonicity property, which allow translating \emph{optimal scheduling} into a mathematical optimization problem, i.e., constraint satisfaction problem (CSP) \cite{CSP}. This is unachievable with the random forest-based models in \cite{Vidur}.
Based on these, we propose new effective preemption and cache replacement policies.
}

\kkim{Our final goal is to save GPU hours in the actual inference serving workloads outside the simulations. This corresponds to our inference system part in Figure \ref{fig:overview}. We integrate/deploy our ideas into \nexttodo{actual systems, including \vLLM,} and prove that these can actually save GPU hours in \emph{online} workloads, where requests may continuously arrive, without performance regression and losing fairness in request serving. 
\kkim{We use \vLLM as the main inference system, and} the monitored performance can be sent to the simulator to calibrate the cost models.
}

\section{Simulator Design}\label{sec:sim_design}



\kkim{This section explains our simulator design: schedulers (Section \ref{subsec:sim_design:scheduler}) and cost models (Section \ref{subsec:sim_design:cost_model}).}

\subsection{Schedulers}\label{subsec:sim_design:scheduler}


\kkim{As mentioned in Section \ref{sec:background}, we focus on preemptive schedulers in open-source online inference systems. 
Their schedulers behave quite similarly, serving requests in first-come-first-serve (FCFS) manner. 
Still, each batch can contain requests in different phases (prefill or decode), and schedulers may prioritize different phases.
We pick two representative schedulers from \vLLM \cite{vLLM} and \Sarathi \cite{SARATHI, SARATHI_SERVE}, each of which prioritize running prefill and decode requests, respectively. \Sarathi further enables \emph{chunked prefill} and \emph{hybrid batching} as explained in Section \ref{subsec:background:related}.
These correspond to deployable schedulers in Figure \ref{fig:overview}.
}

\kkim{For a request having $I$ input tokens and generating $O$ output tokens, its peak number of KVs in the KV cache is $I + O - 1$ considering that the last token does not need to be stored to generate another one.
Since the output length $O$ is unknown in advance, the cache initially allocates $I$ tokens for the request \cite{vLLM} or with some margin. Each new KV generated in the decode step is appended to the cache.
When output sizes are large, high contention can occur among concurrent requests competing for appending their KVs to the cache, resulting in request preemptions.}

\kkim{As noted in Section \ref{subsec:background:related}, most systems adopt newest-request first (NRF) preemption policy. 
In subsequent sections, we show that preempting long requests leads to suboptimal performance, and preempting shortest requests first (SRF) improves performance without performance regression while preserving fairness.
}

\kkim{For hypothetical schedulers, we consider \emph{preemption-free} (PF) schedulers that reserve the peak KV usage ($I + O - 1$) when starting a request, assuming that the output size $O$ is known in priori. An arbitrary preemptive scheduler can be converted into a preemption-free counterpart.}

\subsection{Cost Models}\label{subsec:sim_design:cost_model}

This section explains the cost models for estimating batch execution times.
We primarily consider GPU time as the main overhead as in \cite{Vidur}. CPU overheads such as scheduling can be largely overlapped, e.g., with asynchronous scheduling, leaving GPU computation as the critical path \cite{NanoFlow}.
In databases, this is analogous to first focusing on the query optimization performance in terms of execution time savings, rather than the query optimization time itself, which is out of our scope.

\kkim{Processing a batch of requests consists of feeding the embeddings of tokens to process in requests to the first LLM layer and consecutively feeding the previous layer's outputs as the inputs to the next layer. The final layer's outputs are fed to the softmax function to generate an output token.
}

\begin{figure}[h!]
\centering
\includegraphics[width=1.0\columnwidth]{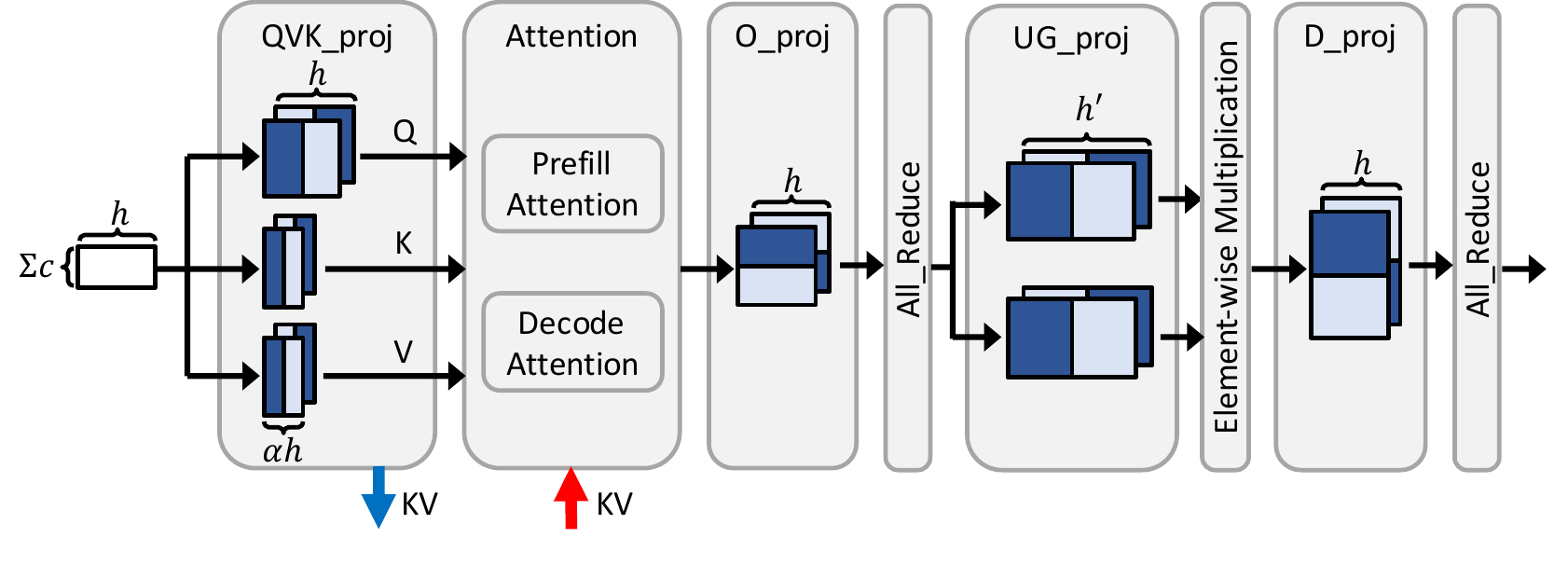}
\caption{\todo{A layer of Transformer architecture. Gray boxes are operators (omitted layernorms, activations, and other operators with negligible overheads). Blue boxes are model weights (matrices), where the arrow located on the left/right indicates input/output of matrix multiplication (matmul). Model weights are partitioned across two GPUs assuming tensor parallelism degree of 2. All\_Reduce adds intermediate results from all GPUs.
QKV\_proj generates KVs and attention reads KVs.
}}\label{fig:transformer_layer}
\end{figure}

\todo{Figure \ref{fig:transformer_layer} shows the intra-layer operators of the Transformer architecture.
For non-attention operators including matrix multiplications (matmuls), all requests' tokens' intermediate results (activation values) are concatenated and processed together, since the operators have distributivity over concatenation, and the overhead of loading the model weights (matrices) can be amortized across the requests in the same batch.
This is represented as $\sum_{r \in \batch} r.c = \sum c$ in Figure \ref{fig:transformer_layer} where $r.c$ denotes the number of tokens to process for a request $r$, and $\batch$ is a batch of requests.
\kkim{$h$ denotes the embedding dimension, the size of embedding for a single token that is input to a layer.}
}

\kkim{Q\_proj, K\_proj, and V\_proj (QKV\_proj) denote the matmuls $f$ in Equation (\ref{eq:attention_formula}). Q\_proj generates the same-size output as the input (by multiplying $(h \times h)$-size matrix), while K\_proj and V\_proj may generate smaller-size outputs. Here, $\alpha \, (\leq 1)$ denotes the degree of grouped query attention (GQA) \cite{GroupedQueryAttention}, and the resulting KV for a token has the size of $2\alpha h$. This is a standard technique to reduce the size of KVs.
The generated KVs are stored in the KV cache.
Note that these matmuls are pre-attention operators, not part of the attention.
}

Regarding the attention operator, the prefill- and decode-attention in Figure \ref{fig:transformer_layer} perform the attention of the same-phase requests together (but using the same kernel implementation) \cite{vLLM}.
\kkim{Attention operators read the KVs of prior tokens specific to each request.
In the standard multi-headed attention \cite{attention}, each attention head has dimension $H = h/N_Q$ where $N_Q$ is the number of query heads.
Since $H$ is the same for K and V, the number of KV heads, $N_{KV}$, is equivalent to $\alpha N_Q$ from $H = \alpha h / N_{KV}$.
}

\kkim{The following operators including three matmuls, O\_proj, UG\_proj, and D\_proj, generate the same-size output ($h$ per token) as the input embedding to the layer and pass it to the next layer.
$h'$ denotes the hidden dimension, typically larger than $h$ \cite{Llama2}.
When tensor parallelism is enabled, model weights are equi-partitioned to multiple GPUs, and All\_Reduce (element-wise) adds the broadcasted outputs from other GPUs.}


\kkim{Now, to build the cost models for intra-layer operators, we start from theoretical cost models \cite{LLMViewer} that are solely based on the LLM and hardware characteristics instead of actual operator execution times.
The LLM determines the number of floating point operations (FLOPs) and amount of data read/write (RW) per intra-layer operator.
}

Table \ref{table:flops_and_data} shows the variables that determine FLOPs and RW.
\kkim{Here, we consider a single request's input $c$ tokens instead of $\sum c$ for simplicity.}
For example, for a matmul between $(c \times h')$ and $(h' \times h)$-size matrices (D\_proj in Figure \ref{fig:transformer_layer}), 
it takes $2ch'h$ FLOPs \cite{NanoFlow}, loading $h'h$ model parameters and $ch'$ inputs, and storing $ch$ outputs. 
Therefore, both FLOPs and RW are linear to $c$, and the bias term captures the cost of loading model weights since both $h'$ and $h$ are fixed.
In case of tensor parallelism, the amount of data to transfer between GPUs in {All\_Reduce} is also linear to $c$.

\begin{table}[]
\renewcommand{\tabcolsep}{1mm}
\caption{\todo{The number of floating point operations (FLOPs) and amount of data read/write (RW) per operator and request. \emph{Others} adds the values of all omitted operators in Figure \ref{fig:transformer_layer}. For brevity, only the request-dependent variables are shown in each cell, which are all linear, e.g., RW for QKV\_proj is $a_1 c + a_2$ for some model/hardware-dependent coefficients $a_1$ and $a_2$.}}
\label{table:flops_and_data}
\begin{tabular}{c|cccc}
\toprule
Operator                  & *\_proj  & Attention                    & All\_Reduce   & Others \\ \hline
FLOPs                     & $c$    & $c^2, m c$             & $c$             & $c$    \\ 
RW           & $c$    & $c^2, m c, c, m$   & $c$         & $c$    \\
\bottomrule
\end{tabular}
\end{table}

For attention, we enable the standard \FlashAttention technique \cite{FlashAttention, FlashAttention_2, FlashAttention_3} and calculate its FLOPs and RW as in \cite{LLMViewer}, which is more complicated than matmuls. If all $B$ requests in a batch have the same number of input tokens $c$ and $m$ KVs to read from the KV cache, and $H$, $N_Q$, and $N_{KV}$ denote the head dimension, number of query heads, number of KV heads, then

\begin{equation}\label{eq:attention_flops}
\begin{split}
    \text{FLOPs} = 4 c (c+m) B H N_Q,
\end{split}
\end{equation}

\begin{equation}\label{eq:attention_data}
\begin{split}
    \text{RW} = & 2 c H N_Q + 2 c (c+m) B N_Q \\ + & 2 \ceil{c/H} (c + m) B H N_{KV}.
\end{split}
\end{equation}

\todo{The hardware (GPUs) determines the FLOPS (computational capacity of GPUs in terms of operation throughput, in FLOPs/s) and bandwidth (in byte/s), we can compute an operator's theoretically optimal execution time as:}

\begin{equation}\label{eq:latency_flops_data}
\begin{split}
    T_{opt} = max\Big(\frac{\text{FLOPs}}{\text{GPU\_FLOPS}}, \frac{\text{RW}}{\text{GPU\_bandwidth}}\Big),
\end{split}
\end{equation}

\noindent assuming complete overlap between computation and data transfer, which leads to discrepancies with practice \kkim{(we show these in Section \ref{subsec:sim_result:batch_compute_memory}).}

\begin{figure}[ht]
    \centering
    \subfigure[Non-attn.]{%
        \includegraphics[width=0.15\textwidth]{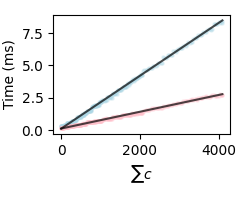}
        \label{fig:regression_other}
    }
    \hfill
    \subfigure[Decode-attn.]{%
        \includegraphics[width=0.15\textwidth]{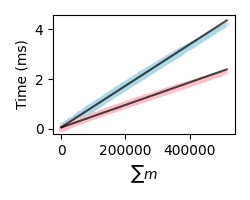}
        \label{fig:regression_decode}
    }
    \hfill
    \subfigure[Prefill-attn.]{%
        \includegraphics[width=0.15\textwidth]{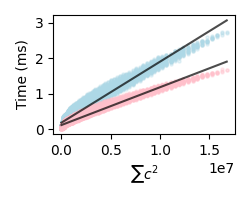}
        \label{fig:regression_prefill}
    }

    \caption{Intra-layer operator execution times measured for the Llama-2-7B model on one A100 (blue) and \todo{H100 (red)}. Non-attn. adds all non-attention operators. Black lines are single-variable linear regressions with $R^2$ scores over 0.96. $\sum c$, $\sum m$, and $\sum c^2$ denote the total number of tokens to process, KVs to read, and quadratic sum of tokens in a batch.}
    \label{fig:regression}
\end{figure}

Still, the theory well explains the linearity of actual operator times in Figure \ref{fig:regression}, where x-axes show the representative variables from Table \ref{table:flops_and_data}.
Note that non-attention operators largely depend on the number of tokens to process, $c$, as both FLOPs and RW are linear to $c$. 
The decode-attention is bottlenecked by reading KVs from GPU memory, whose time is therefore correlated with $m$, the number of KVs to read from the cache.
The prefill-attention time scales with $c^2$ due to the quadratic complexity of attention, but more importantly, the quadratic amount of \emph{data transfer} \kkim{(we show in Section \ref{subsec:sim_result:batch_compute_memory}).}
Given an LLM and GPU setup, we train linear cost models, the coefficients of all variables in Table \ref{table:flops_and_data} (theory), using actual operator times as labels (practice), over diverse number of tokens to process $c$, number of KVs to read $m$, and batch size $B$.

To predict batch execution times, we sum the predicted execution times of all intra-layer operators and multiply by the number of layers.
In our comparison with actual end-to-end workload runtimes \kkim{in Section \ref{sec:sim_result}}, 
the average and maximum relative error of the predictions is 6\% and 12\%, showing that simple linear models are effective enough to \kkim{capture inference performance.}


Note that our focus is not on building near-perfect cost models, but to \kkim{1) save GPU hours in the development and analysis phase} and \kkim{2)} explain the phenomena in LLM inference \kkim{and derive meaningful insights to develop new techniques in actual inference systems.}
\kkim{In contrast, \Vidur \cite{Vidur} focuses on accurate predictions, reporting the maximum error of 9\% using random-forest models. Its cost models are not theoretically grounded, nor closed-form or monotonic, thus it is unable to perform our theoretical analysis in Section \ref{sec:theory} including the CSP.
Furthermore, its goal is to find the best system configurations for specific workloads with exhaustive search, not to develop new techniques.
}

\section{Simulator Results}\label{sec:sim_result}








\kkim{This section shows that the simulation can save GPU hours in the development and analysis phase in LLM inference systems, proving it as an effective tool. By comparing with actual inference performance, we also address research questions that have received limited attention or generated misbelieved answers.}

\kkim{After our setup in Section \ref{subsec:sim_result:setup}, Sections \ref{subsec:sim_result:batch_compute_memory}-\ref{subsec:sim_result:batch_swap_recompute} are based on theoretical cost models and actual measurements, and Sections \ref{subsec:sim_result:preemptive_simulation}-\ref{subsec:sim_result:increasing_M} are based on hybrid cost models (learned coefficients on top of theoretical models), and Section \ref{subsec:sim_result:preemptive_actual} is based on actual measurements to confirm the simulation results.}



\subsection{Setup}\label{subsec:sim_result:setup}

\noindent \underline{\textbf{Model and Hardware.}}
We follow the default configuration of \cite{Vidur}, using the Llama-2-7B model on an \todo{A100 or H100 GPU with 80GB memory} and \kkim{the KV cache size to 100K} as default.
{In Section \ref{sec:deploy} on real-world systems, we scale up to Llama-3-70B model with context size $S$ of 128K, on four GPUs to show that the insights found in this section generalize to other hardware-model configurations.}
{We use 4 GPU hours to calibrate our cost models.}

\noindent \underline{\textbf{Metrics.}} 
\todo{We mainly use the end-to-end latency and time-per-output-token (TPOT) that measure the system-side and user-side inference performance. Tokens-per-second (TPS), the number of generated tokens divided by latency, 
and time-to-first-token (TTFT), the delay to generate the first token, follow similar trends with the latency. The latency is critical for measuring GPU hour savings \kkim{in online serving}, and data analytics workloads using LLMs \cite{LOTUS} consider latency as the primary metric \kkim{as in the database OLAP workloads}. TPOT is an indicator for service-level objective (SLO) attainment similarly to constant-delay algorithms in databases \cite{ConstantDelay}; larger TPOT indicates less interactivity in user experience.
}

\subsection{What Makes a Batch Compute-Bound?}\label{subsec:sim_result:batch_compute_memory}







\todo{Previous studies \cite{SARATHI, SARATHI_SERVE, DeepSpeed} have shown that matmuls (`*\_proj' in Figure \ref{fig:transformer_layer}) take the majority of the runtime for both prefill- and decode-batches despite the quadratic complexity of the attention computation, for diverse $(c, m)$ values \kkim{($c$: number of tokens to process, $m$: number of KVs to read)} and batch sizes. The quadratic computation is what has been believed to make prefill-attention compute-bound and decode-attention memory-bound \cite{NanoFlow}, determining the resource utilization of a batch.}

\begin{figure}[h!]
\centering
\includegraphics[width=0.87\columnwidth]{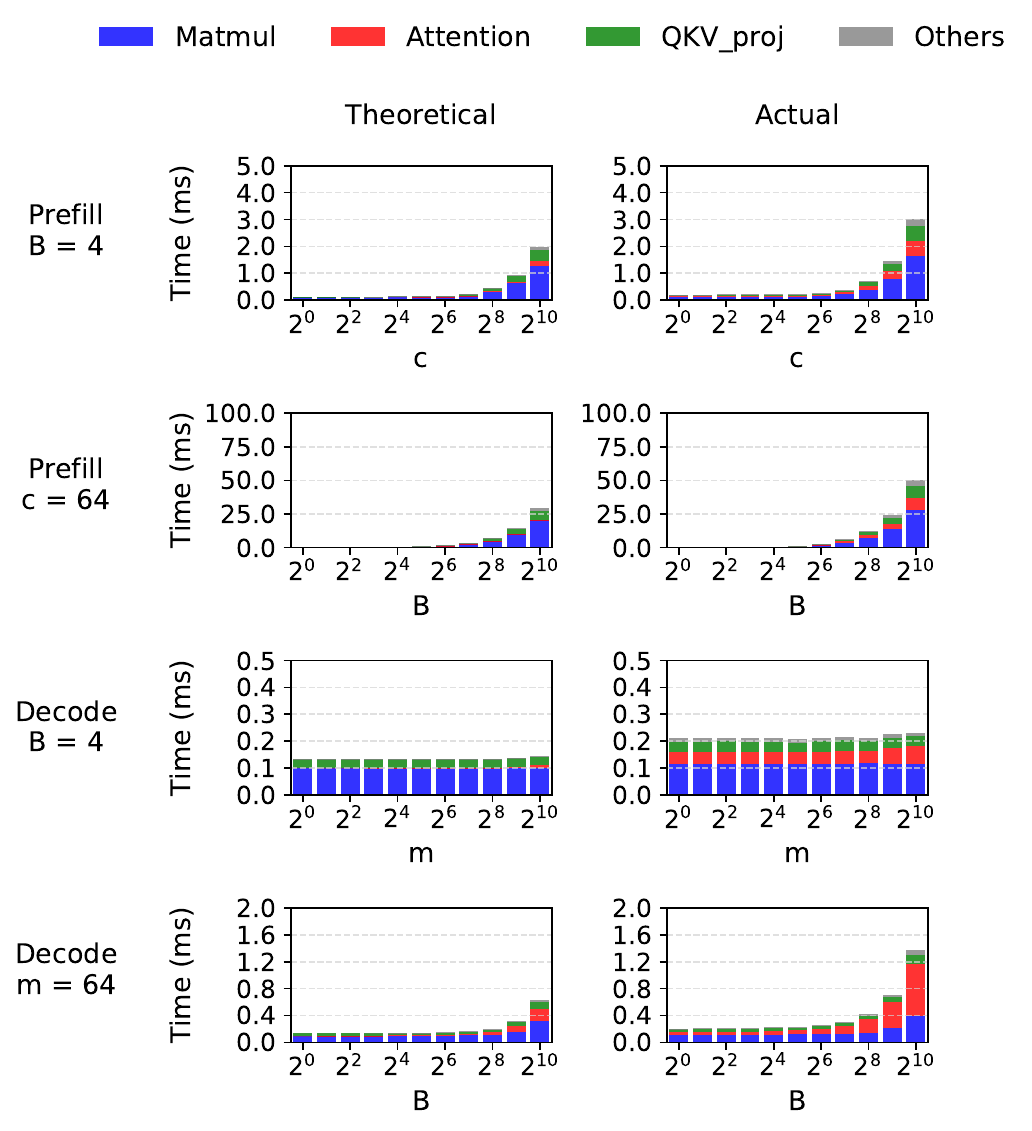}
\vspace*{-0.2cm}
\caption{\todo{\kkim{Theoretical and actual} time breakdowns for prefill and decode batches on H100. Matmul includes all `*\_proj' operators in Figure \ref{fig:transformer_layer} except QKV\_proj. All $B$ requests in a batch have the same $c$ (number of tokens to process) and $m$ (number of KVs to read) values.}}\label{fig:operator_time_breakdown}
\end{figure}

\todo{However, we show that there are misbelieves. We first breakdown and compare the operator times in Figure \ref{fig:operator_time_breakdown}: the theoretical batch execution times with Equation (\ref{eq:latency_flops_data}) and actual times. It also shows that matmuls are the major bottleneck. However, while the complexity of prefill-attention is quadratic to $c$ (Table \ref{table:flops_and_data}), the attention becomes the actual bottleneck for decodes with $c=1$, not prefills, for larger batch sizes and $m$.
Furthermore, attentions have a larger gap between theoretical times and actual ones than the other operators.
}

\begin{figure}[h!]
\centering
\includegraphics[width=0.9\columnwidth]{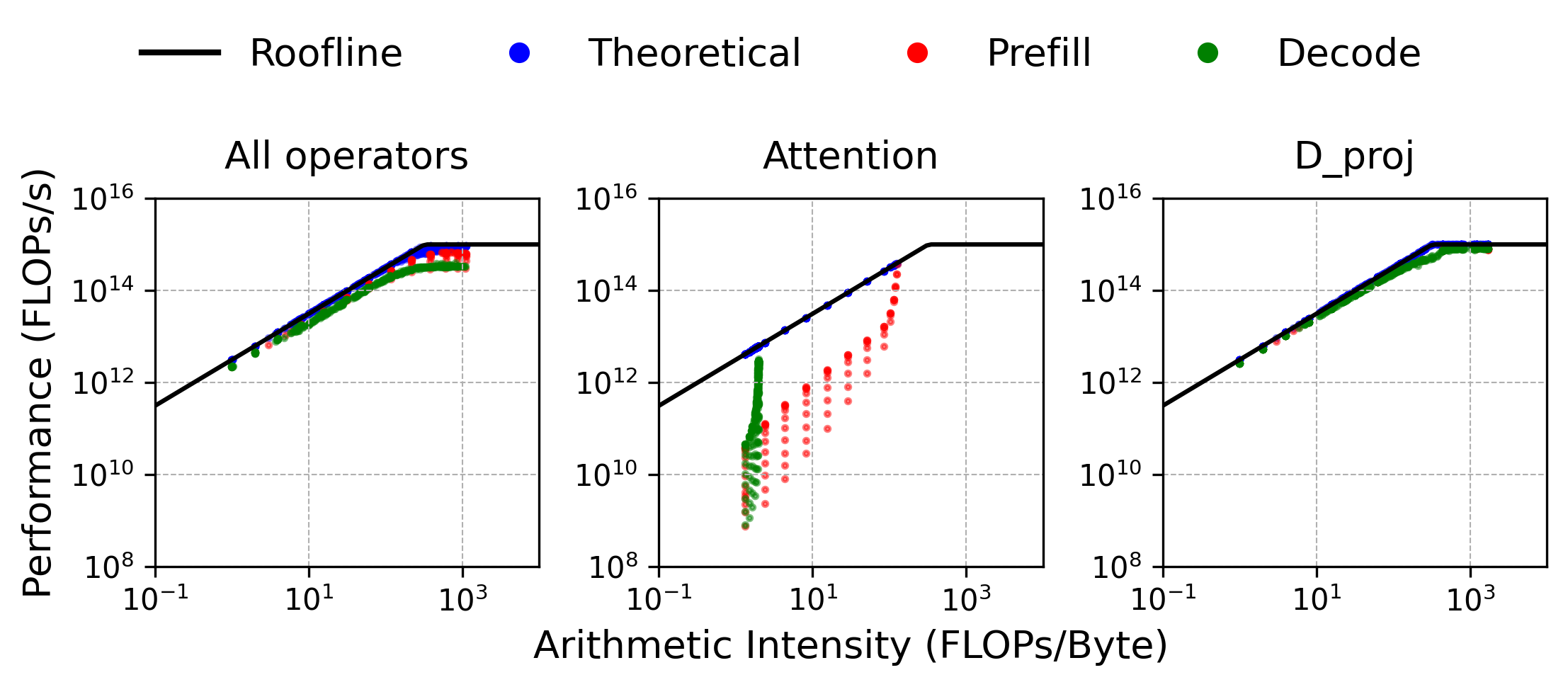}
\vspace*{-0.2cm}
\caption{\todo{Roofline analysis on H100. Each dot corresponds to a batch. \emph{Prefill} and \emph{Decode} are based on actual time measurements, \emph{Theoretical} results from Equation (\ref{eq:latency_flops_data}). Other `*\_proj' exhibit similar results with D\_proj.}}\label{fig:roofline}
\vspace*{-0.2cm}
\end{figure}

\todo{To clarify these, we adapt the roofline analysis \cite{LLMViewer} in Figure \ref{fig:roofline}. The roofline shows the hardware characteristic; it is a union of the left memory-bound region with increasing performance bound (slope is the GPU bandwidth) and the right compute-bound region with fixed performance bound (of GPU FLOPS) with a turning point in between.
An operator's intensity (FLOPs/RW) determines its compute-/memory-boundness, and performance is the achieved throughput of FLOPs.
\emph{Note that all attention points are memory-bound, even for the prefills.} Hence, the attention time scaling with $c^2$ in Figure \ref{fig:regression_prefill} is due to the quadratic data transfer cost in Equation (\ref{eq:attention_data}), not its computation.
Attentions are also distant from the roofline, indicating the severe under-utilization of GPU bandwidth due to interleaved (not overlapped) computations. This explains the gap with theoretical times in Figure \ref{fig:operator_time_breakdown}, \kkim{and justifies fitting the cost models with actual times in Section \ref{subsec:sim_design:cost_model}.}
}

\noindent \underline{\textbf{Remark.}} \todo{Attentions are memory-bound. Only matmuls can be compute-bound, when the number of tokens to process, $c$, is large enough to surpass the cost of loading fixed-size model weights. The whole batch can be compute-bound when 1) matmuls are compute-bound, and 2) the number of KVs to read, $m$, is not large enough to make the attention dominate the batch time. Decode-batches can also be compute-bound (Figure \ref{fig:roofline}).
}


\todo{Because of the sequential nature of intra-layer operators, we need to make every operator balance compute and memory to make the whole batch balanced, more than simply batching prefill- and decode-requests together as in \cite{SARATHI_SERVE}.
The challenge is in achieving the balance for attentions. In Figure \ref{fig:roofline}, prefill-/decode-attention batches converge to certain points close to the roofline, as $c, m$, and $B$ (batch size) increase. From Equations (\ref{eq:attention_flops}) and (\ref{eq:attention_data}), the intensity (FLOPs/BW) converges to $2 / (1/H + \ceil{c/H} N_{KV} / (c N_Q))$. For the Llama-2-7B model, $H = 128$ (head dimension) and $N_{Q} = N_{KV} = 32$ (number of query and KV heads). For prefills with large $c$, the intensity becomes $2 / (1/128 + 1/128) = 128$. For decodes with $c=1$, it becomes $2 / (1/128 + 1) \approx 2$. Therefore, to balance attentions, either the model should have higher $H$ and $N_Q/N_{KV}$ values or the GPU bandwidth should be higher to push the turning point leftwards.
}







\vspace*{-0.2cm}

\subsection{Should We Recompute or Swap KVs?}\label{subsec:sim_result:batch_swap_recompute}

In case of preemptions, when evicting KVs from GPU memory and refilling them, two representative options are: (1) recomputing KVs and (2) swapping in the offloaded KVs from DRAM.
Without specialized techniques \kkim{(Section \ref{sec:background})}, (2) is largely inefficient than (1) due to the low PCIe bandwidth \cite{vLLM, InfiniGen}.

\begin{figure}[h!]
\centering
\includegraphics[width=0.93\columnwidth]{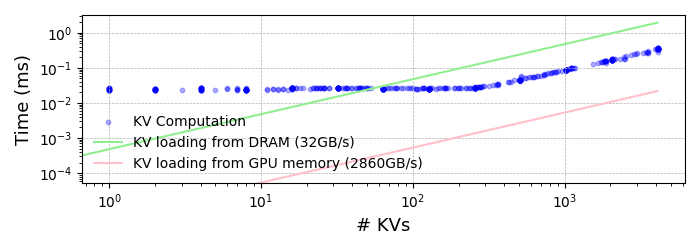}
\vspace*{-0.5cm}
\caption{\kkim{KV refill times by recomputation (blue, actual measurements) and loading from DRAM (green, theoretical). For comparison, loading from GPU memory (KV cache) is also shown (red, theoretical).}}\label{fig:KV_swap_recompute}
\end{figure}

Figure \ref{fig:KV_swap_recompute} shows the measured points for (1) and \kkim{theoretically} optimal performance for (2) over varying number of KVs. When \# KVs is large, (1) is more efficient than (2). But when \# KVs is small, (2) can be more efficient since (1) suffers from a fixed cost of loading model weights. Still, the turning point is small (less than 100 KVs) compared to the full KV cache size (100K KVs).
The cost for loading KVs from GPU memory shows the advantage of using KV cache than recomputing KVs every time.

\noindent \underline{\textbf{Remark.}} \todo{While the recomputation is more efficient than swapping in general, swapping could improve efficiency with future PCIe generations, in small-cache environments such as end-devices, or when sharing GPU memory with other applications (multi-tenancy). As shown in \kkim{Section \ref{subsec:sim_result:batch_compute_memory},} the bandwidth is crucial in LLM inference efficiency.}

\subsection{Preemptive Schedulers (Simulation)}\label{subsec:sim_result:preemptive_simulation}

\todo{Now we consider \kkim{preemptive schedulers in open-source inference systems} and focus on the effect of preemption \kkim{on inference performance}. 
We show the results on A100 where the results on H100 are similar (while faster), and use KV recomputation to refill KVs in case of preemption. 
Note that previous analyses \cite{vLLM, SARATHI, SARATHI_SERVE, DeepSpeed} have largely focused either on the single-batch cases or simple trade-offs between latency and TPOT for multi-batch cases without focusing on preemption.
}


\noindent \underline{\textbf{Workloads.}} 
\kkim{For the analysis purpose, long, complex workloads with varying input and output lengths make it difficult to understand the behaviors of schedulers and discover new insights for improving them.}
Hence, to simply the analysis, here we use \emph{offline} workloads that all requests arrive at time zero, with the same \kkim{input size} $I$ and \kkim{output size} $O$.
\kkim{Still, we later remove this assumption and use \emph{online} workloads (continuously arriving requests) and heterogeneous $I$ and $O$ values in Section \ref{sec:deploy}.}
Please refer to our technical paper \cite{kim2024faster} that simulates and compares the schedulers prioritizing requests based on their lengths under heterogeneous requests.
We vary the $I$ and $O$ values from 1 to 1024 to cover a wide range of workloads where each request cannot be longer than Llama-2-7B's context length, 4096. This range covers short-answer questions to long-text generations\footnote{A real-world \todo{chat} workload \cite{ShareGPT} shows an average input length of 70 and output length of 215. \todo{Data analytics workloads have shorter outputs, e.g., on average less than 10 tokens for table question answering \cite{WikiTableQuestions} and 50 for text-to-SQL \cite{BIRD}.}}.
\kkim{We later use longer requests in Section \ref{sec:deploy}.}
\kkim{We fix the number of requests to 1024 to represent high-contention scenarios.}

\noindent \underline{\textbf{Schedulers.}} 
\kkim{As noted in Section \ref{subsec:sim_design:scheduler},}
we employ the schedulers of \vLLM \cite{vLLM} and \Sarathi \cite{SARATHI_SERVE} as \kkim{representative preemptive schedulers that show the best inference performance in \cite{Vidur}.}
Table \ref{table:schedulers} shows the schedulers we compare, including the variants of the two but excluding the ones that either performed similarly or worse in our experiments.

\begin{table}[]
\renewcommand{\tabcolsep}{1mm}
\caption{Preemptive schedulers used. $C$ denotes the max number of tokens to process per batch.
}
\vspace*{-0.2cm}
\label{table:schedulers}
\begin{tabular}{c|cccc}
\toprule
Scheduler           & Priority & Hybrid Batch & Chunked Prefill & $C$  \\ \hline
\vLLM               & Prefill  & \xmark            & \xmark               & 4096 \\ 
\Sarathi            & Decode   & \cmark            & \cmark               & 512  \\ \hline
\SarathiPC          & Decode   & \cmark            & \cmark               & 4096  \\
\SarathiNOCP        & Decode   & \cmark            & \xmark               & 4096  \\
\vLLMHY             & Prefill  & \cmark            & \xmark               & 4096  
\\ \bottomrule
\end{tabular}
\end{table}

\begin{figure}[h!]
\centering
\includegraphics[width=1.0\columnwidth]{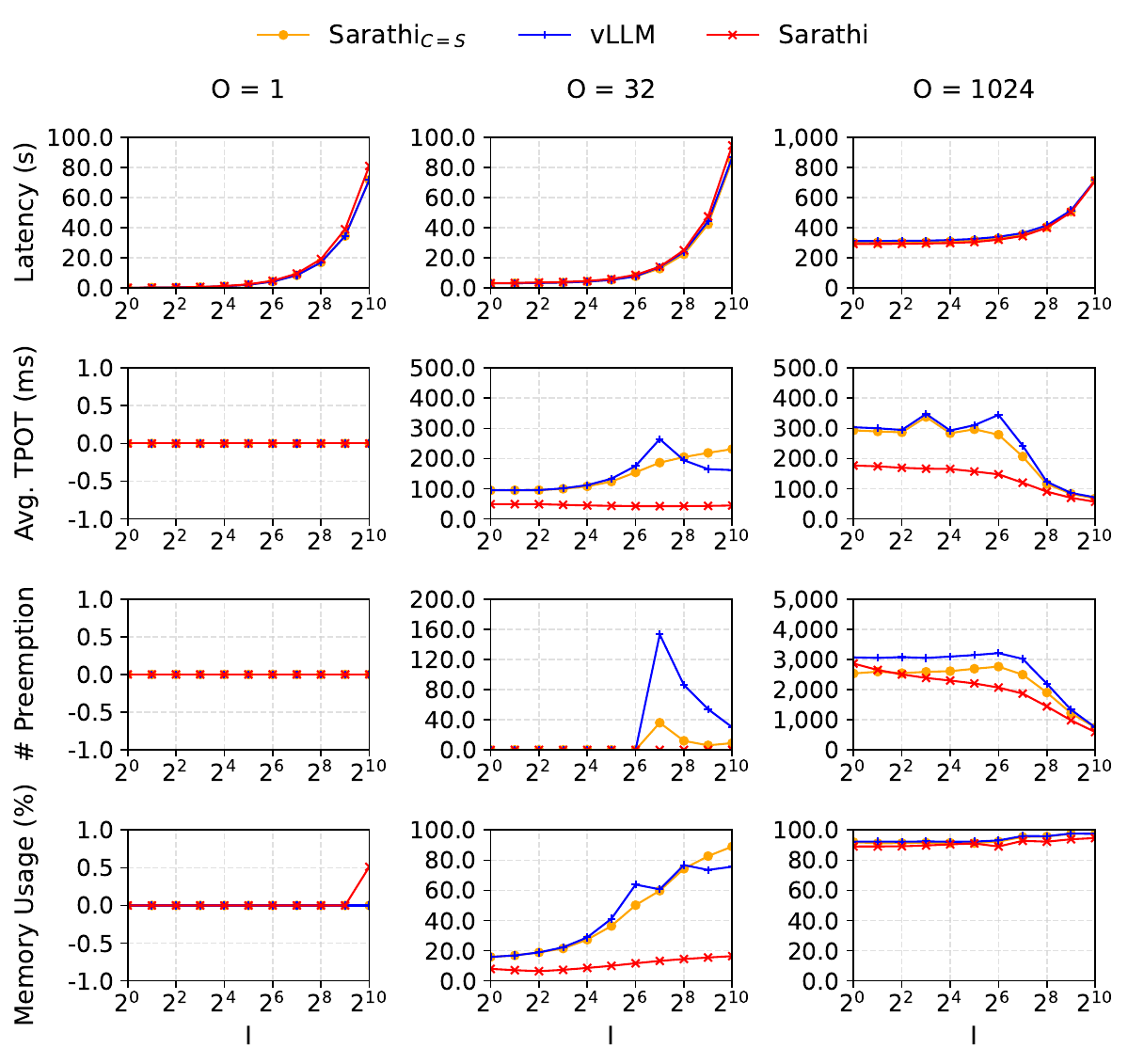}
\vspace*{-0.5cm}
\caption{Results on varying input size $I$ and output size $O$ of requests. \todo{Memory usage indicates KV cache usage.}}\label{fig:exp_large_B}
\end{figure}

We group the schedulers in Table \ref{table:schedulers} based on their performance into: 
1) \Sarathi, 2) \SarathiPC and \SarathiNOCP, and 3) \vLLM and \vLLMHY. 
Figure \ref{fig:exp_large_B} shows the results for \Sarathi, \SarathiPC, and \vLLM as representatives.

\noindent \underline{\textbf{Latency and TPOT.}}
Overall, as the input size $I$ and output size $O$ increase, latency increases across all schedulers. 
\vLLM shows the lowest latency 
by prioritizing prefill requests and batch processing decodes in parallel, resulting in large batch sizes, except when high $O$ values lead to frequent preemptions. Preemption rates increase with $O$ because each request competes to keep more tokens in the KV cache. 
\Sarathi generally has the highest latency (up to \todo{13\%} higher than \vLLM) but achieves a stable TPOT (up to \todo{5.3x} lower than \vLLM) due to balanced handling of prefill and decode phases through hybrid batching and a smaller batch-wise token limit $C$. \todo{This shows the trade-off between latency and TPOT as in \cite{SARATHI, SARATHI_SERVE}.}
\SarathiPC 
resembles \vLLM 
as it processes up to the same \todo{$C$} prefill tokens per batch, matching \vLLM's prefill speed by managing up to $C/I$ new running requests per batch. 
\todo{TPOT increases with $C$ due to larger number of tokens to process, larger batches from higher prefill speed, and more KVs read in subsequent decode steps.}

An interesting point is that TPOT decreases beyond a certain value of $I$. \mlsys{For large $O$ (e.g., 1024), frequent preemptions cause refills to dominate runtime, while larger $I$ limits batch sizes and reduces preemptions. This reduces time intervals between token generations.}
For small $O$, refills have a smaller impact.

\noindent \underline{\textbf{Preemption.}} The key distinction between the input size $I$ and output size $O$ in terms of their effect on preemption is that $I$ represents the immediate memory reserved, whereas $O$ determines the peak memory usage after approximately $\Omega(O)$ batches have been processed. Consequently, schedulers that only consider $I$ but not $O$ risk overloading the system by batching requests with their long-term memory demands underestimated. As $O$ increases, this can lead to a significant increase in the number of preemptions.

\noindent \underline{\textbf{Memory (KV Cache) Usage.}}
Because \Sarathi gradually adds running requests, it maintains stable memory consumption for KVs. 
However, for $O = 1024$, \Sarathi experiences high memory demand, occupying more than 90\% of the KV cache alike other schedulers.

\noindent \underline{\textbf{Remark.}}
Schedulers face a basic \todo{latency}-TPOT trade-off \cite{SARATHI, SARATHI_SERVE}. \Sarathi maintains a balanced \todo{latency} and TPOT across varying $I$ and $O$ values. Other schedulers excel with moderate values but encounter preemption spikes under high memory demands due to aggressive batching. \Sarathi mitigates this with a smaller max number of tokens per batch, $C$. 
High TPOT results primarily from preemption, with batch size and KV load as secondary factors. Preemption increases with larger $O$ values, while $I$ acts as both a limiting factor (as new running requests are bounded by $C/I$) and an increasing factor (by raising memory demands).

\subsection{How Good Is It to Avoid Preemption? (Simulation)}\label{subsec:sim_result:how_good_to_avoid_evictions}

\begin{figure}[h!]
\centering
\includegraphics[width=1.0\columnwidth]{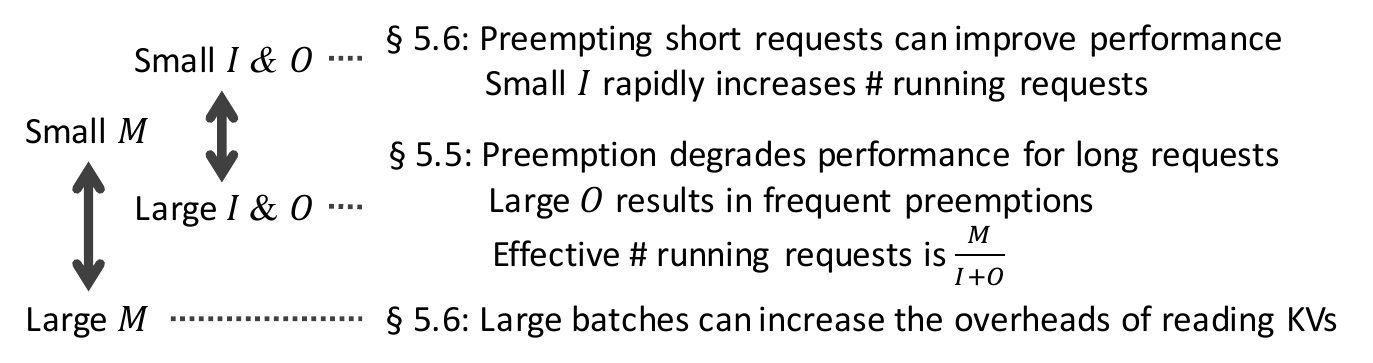}
\vspace*{-0.2cm}
\caption{An overview of key insights in upcoming sections. $I$ and $O$ denote the input and output size of requests, and $M$ denotes the KV cache size.}\label{fig:experiment}
\end{figure}

With a basic understanding of the factors affecting performance, we explore some key questions. Figure \ref{fig:experiment} provides a high-level view. We begin by comparing the schedulers in Table \ref{table:schedulers} to their hypothetical, preemption-free (PF) counterparts in Table \ref{table:schedulers}.
We use the output size $O$ of 1024, a scenario with frequent preemptions, since in other cases, PF schedulers perform similarly to their original \mlsys{(non-PF)} counterparts.

\begin{figure}[h!]
\centering
\includegraphics[width=0.7\columnwidth]{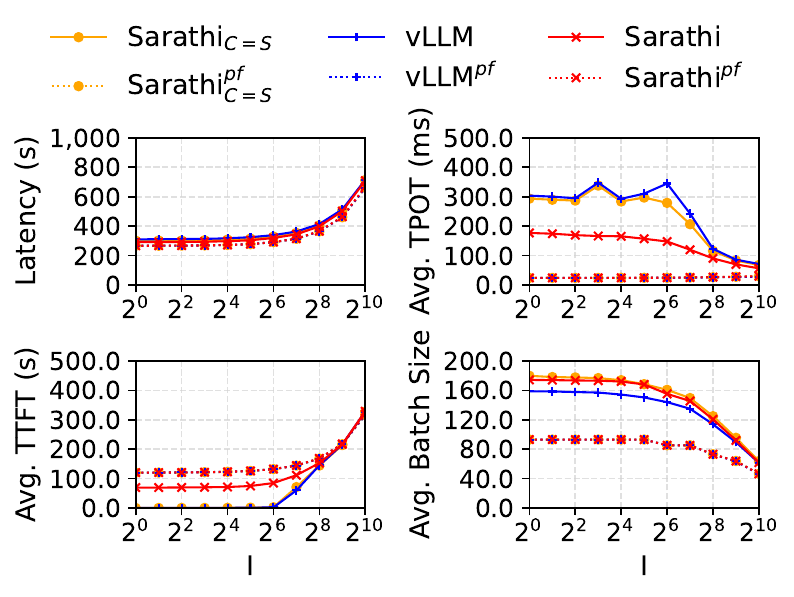}
\vspace*{-0.2cm}
\caption{Results on varying input size $I$ and the fixed output size $O$ of 1024 of requests. Hypothetical, preemption-free (PF) schedulers are indicated by the {\textit{pf}} superscript in the labels.
}\label{fig:exp_bin_packing}
\end{figure}

As illustrated in Figure \ref{fig:exp_bin_packing}, \mlsys{PF} schedulers generally achieve better system performance, with \todo{latency reduction} over their \mlsys{non-PF} counterparts reaching \todo{up to 17\%, 10\%, and 14\%} for \vLLM, \Sarathi, and \SarathiPC, \todo{since there is no refill overhead.}
However, PF schedulers exhibit higher TTFT since they \mlsys{try to reserve more memory in initiating requests, waiting more for the current} running requests to complete and release their KVs. This TTFT increase is substantial -- \todo{up to 1000x for \vLLM and \SarathiPC, and 1.7x for \Sarathi} -- but is offset by lower TPOT, with reductions of \todo{up to 13x for \vLLM and \SarathiPC, and 7.4x for \Sarathi}.

\noindent \underline{\textbf{Remark.}} \todo{There is a clear trade-off between TTFT and TPOT \cite{SARATHI},}
and it is crucial to limit the number of running requests by considering the size of KV cache, $M$, and the memory demands of requests.
The \emph{effective} \todo{batch} size can be approximated as $\frac{M}{I+O}$, as supported by Figure \ref{fig:exp_bin_packing}, where PF schedulers achieve an average \todo{batch} size close to $\frac{100K}{1+1024} \approx 98$ for $I = 1$ and $\frac{100K}{1024+1024} \approx 49$ for $I = 1024$.

\subsection{Is Increasing the KV Cache Size a Silver Bullet? (Simulation)}\label{subsec:sim_result:increasing_M}



\begin{figure}[h!]
\centering
\includegraphics[width=1.0\columnwidth]{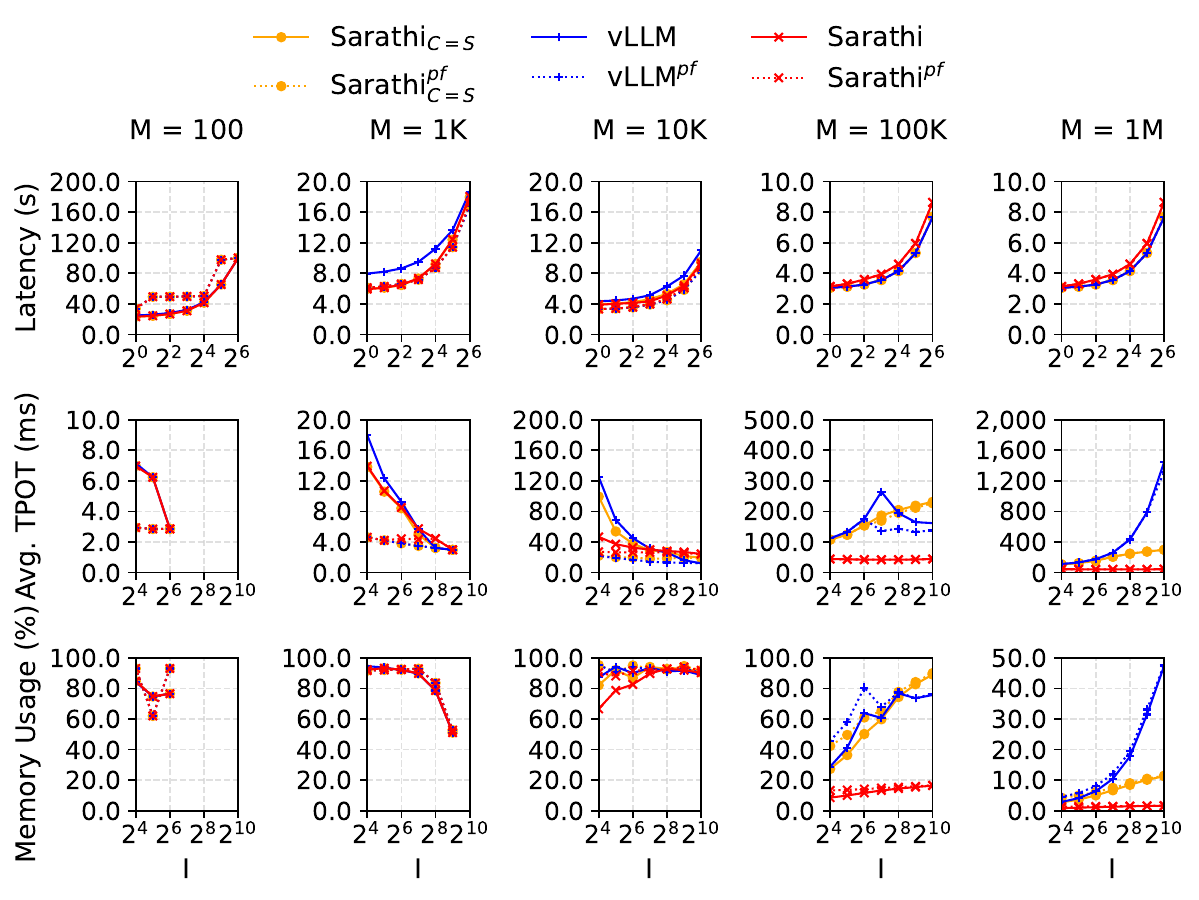}
\vspace*{-0.2cm}
\caption{Results on varying input size $I$ and the fixed output size $O$ of 32, under varying KV cache size, $M$. The x-axes are cropped to display only the areas of interest.}\label{fig:varying_M}
\end{figure}

To avoid preemption and maximize the effective \todo{batch} size, one might wonder if simply increasing the KV cache size, $M$, to a sufficiently large value could solve all issues, \mlsys{possibly with multiple or future-generation GPUs}. To explore this, we vary $M$ from 100 to 1M, testing under different memory contention levels and hardware-model configurations to simulate lower memory budgets. Figure \ref{fig:varying_M} shows results for the output size $O$ of 32.

Interestingly, when the cache size $M$ is small, PF schedulers actually show \todo{higher latency} compared to non-PF ones, as they wait too long for running requests to release memory. \emph{In this context, preemption can rather reduce latency.} 
As shown in Figure \ref{fig:varying_M}, preemption \todo{reduces latency} by \todo{up to 1.9x and 2x for \vLLM and \Sarathi}, under $M$ = 100, and by \todo{1.3x and 1.1x} under $M$ = 1K, mainly due to reduced TTFT.
However, again these gains come with higher TPOT as a trade-off.

\noindent \underline{\textbf{Remark.}}
\todo{Preempting short requests can reduce latency by 2x compared to being too conservative to avoid any preemption under high memory contention, which we \kkim{first evaluate in Section \ref{subsec:sim_result:preemptive_actual}} then theoretically prove in Section \ref{sec:theory}.}
The results also highlight that increasing the cache size alone is not a complete solution.
Enhanced memory bandwidth is essential to minimize the cost of reading KVs \todo{in memory-intensive decodes as well as the quadratic amount of data transfer in prefills} (Section \ref{subsec:sim_result:batch_compute_memory}).
In addition, Figure \ref{fig:varying_M} shows that under a large cache size (1M), \Sarathi and \SarathiPC utilize less than 20\% of KV cache on average, even when processing numerous long requests, highlighting potential underutilization.

\subsection{Results on a Real Inference System (Actual)}\label{subsec:sim_result:preemptive_actual}

\begin{figure}[h!]
\centering
\includegraphics[width=1.0\columnwidth]{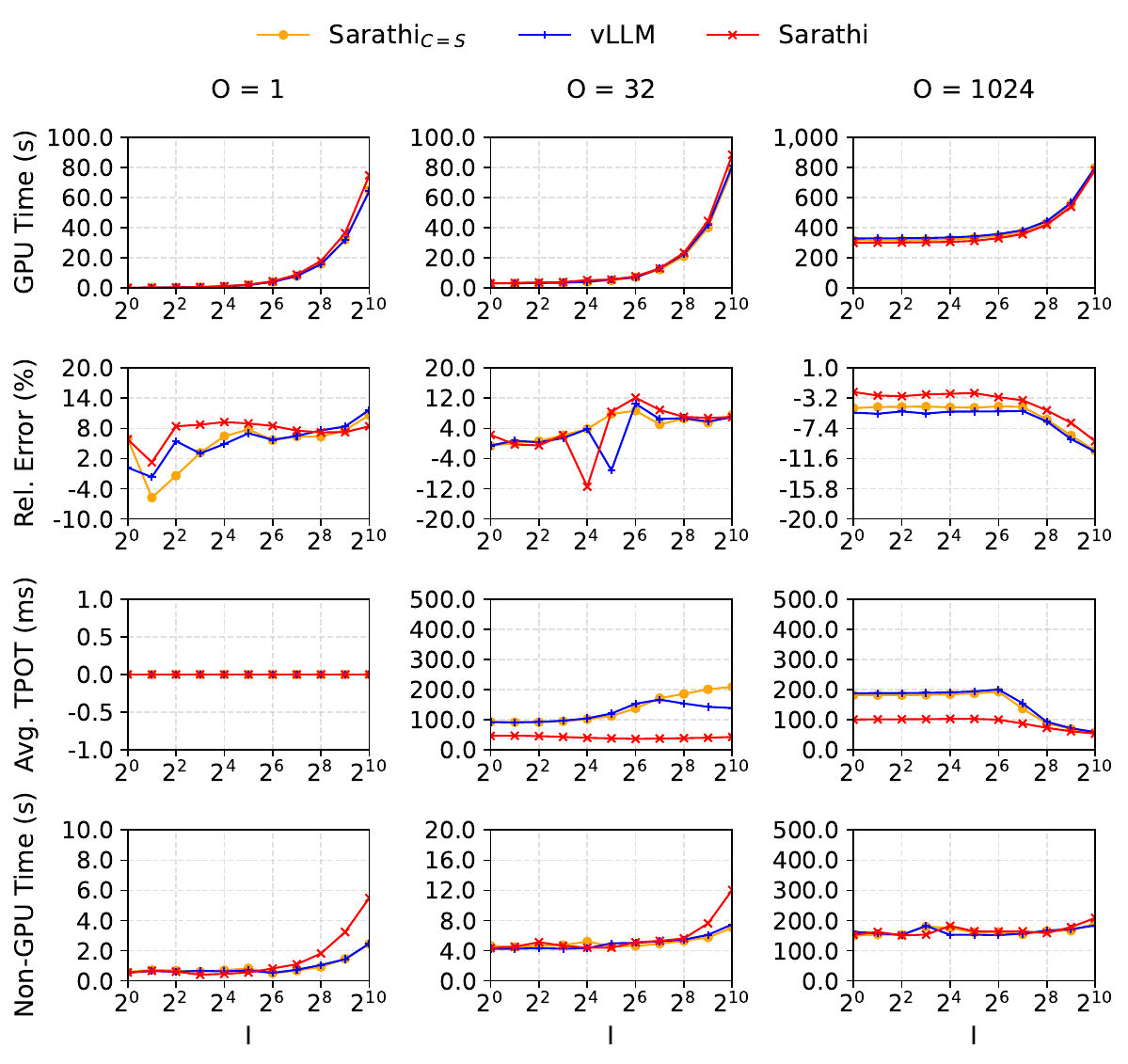}
\caption{Results of running the schedulers in Figure \ref{fig:exp_large_B} on \vLLMS.
GPU time represents the duration for GPU-based operations, including matmuls and attentions, while non-GPU time covers scheduling and token sampling. The second row indicates the relative error of the latency in Figure \ref{fig:exp_large_B} compared to the GPU time. 
}\label{fig:real_exp_large_B}
\end{figure}

\begin{figure}[h!]
\centering
\includegraphics[width=1.0\columnwidth]{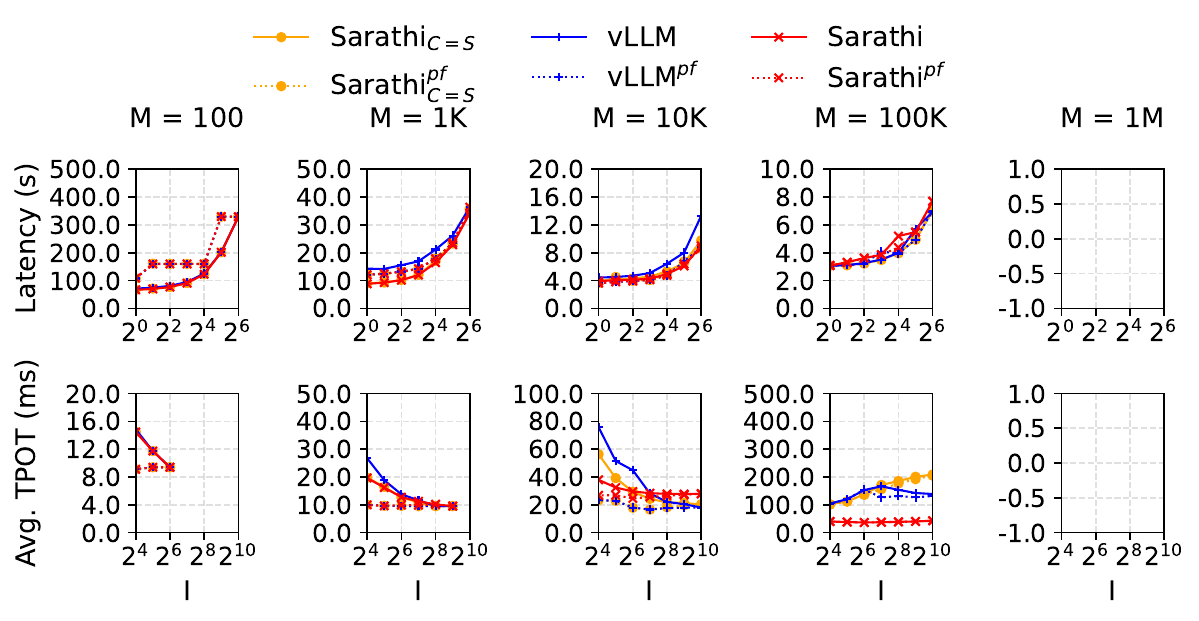}
\vspace*{-0.2cm}
\caption{Results of running the schedulers in Figure \ref{fig:varying_M} on \vLLMS. No results are provided for the cache size $M$ of 1M since the maximum $M$ available on the physical GPU for the Llama-2-7B model is below 1M.}\label{fig:real_varying_M}
\end{figure}

This section presents results from running the experiments in Sections \ref{subsec:sim_result:preemptive_simulation} and \ref{subsec:sim_result:increasing_M} on a real LLM inference system (Figures \ref{fig:real_exp_large_B} and \ref{fig:real_varying_M}). We use \vLLM \cite{vLLM} v0.6.3 and \kkim{run the generated schedules from the simulator. Later in Section \ref{sec:deploy}, we integrate our new preemption and cache replacement policy in the actual system.} 
For clarity, here \vLLM refers to the scheduler as before, while \vLLMS denotes the inference system.


Figure \ref{fig:real_exp_large_B} shows similar results with Figure \ref{fig:exp_large_B}, \todo{with the average and maximum relative error of 6\% and 12\% for latency.
}
\mlsys{Non-GPU time mainly consists of CPU-based scheduling and token sampling, where we leave optimizing them as an orthogonal future work. Recent inference systems and techniques, e.g., asynchronous scheduling \cite{NanoFlow} mentioned in Section \ref{subsec:sim_design:cost_model}, allow effective overlapping of CPU and GPU operations.}

Figure \ref{fig:real_varying_M} shows results for varying cache size $M$. 
Consistent with Section \ref{subsec:sim_result:increasing_M}, preemption \kkim{decreases latency} by \todo{up to 2.3x} for small $M$ values of 100 and 1K, and \kkim{increases latency} by \todo{up to 1.5x} for larger $M$ values.



\noindent \underline{\textbf{Remark.}}
\kkim{The results on an actual inference system confirm our primary focus, that simulations can be an effective tool to save GPU hours in the development and analysis phase of inference systems, while still offering valuable insights.
We re-emphasize that our focus is not on providing near-perfect simulations (e.g., less than 1\% relative error) nor repeating existing findings (e.g., latency-TPOT trade-off), but on investigating new opportunities (e.g., preemptions) to develop better mechanisms in actual inference systems, ultimately reducing GPU hours in online serving.
In subsequent sections, we ground our ideas in theory and then show how they deliver robust empirical results.
}


\section{5-Minute Rule for LLM Inference}\label{sec:five_minute_rule}

This section provides an interesting connection between databases and LLM inference from that they both maintain caches (or buffers). 
\kkim{In Section \ref{subsec:sim_result:batch_swap_recompute}, we have seen that it is more efficient to keep KVs in the KV cache than recomputing (or loading from DRAM), in terms of latency.
However, it is unknown whether its also \emph{cost-effective} in terms of GPU maintenance costs in on-demand cloud GPU environments and LLM services like ChatGPT.
}
That is, what is the break-even interval $T_{break}$ for keeping the KVs in the GPU memory to be cost-effective, \kkim{that if a KV is re-accessed after this interval, then it is better to discard it and recompute it later?}


To answer this, we apply the five-minute rule \cite{FiveMinuteRule} from the database domain, originally proposed for computing the break-even interval at which it becomes cost-effective to keep a data page in memory rather than fetching it from disk each time. The original formula is



\vspace*{-0.2cm}
\begin{equation}\label{eq:five_minute_original}
\begin{split}
    T_{break} & = \frac{\text{\# pages per MB of RAM}}{\text{\# accesses per second per disk}} \\ & \times \frac{\text{\$ per disk}}{\text{\$ per MB of RAM}},
\end{split}
\end{equation}

\noindent \todo{when the price for cache memory (\$/page/s) matches the savings in disk accesses per second (\$/access/s).
Similarly, in our case, the price for KV cache memory (\$/KV/s) should match the savings in KV recomputation per second (\$/recomputation/s), where the price terms cancel (both measured in GPU-seconds):
}

\vspace*{-0.2cm}
\begin{equation}\label{eq:five_minute_llm}
\begin{split}
    & T_{break} \\
    & = \frac{\text{\# KVs per MB}}{\text{\# recomputations per second}} \times \frac{\text{KV cache size}}{\text{\# KVs per MB}} \\ & = t^1_{recom} \times M.
\end{split}
\end{equation}

\noindent \todo{Here, $M$ is the KV cache size and $t^1_{recom}$ is the time for recomputing 1 KV. For $m$ KVs of a request, it becomes $t^m_{recom} \frac{M}{m}$ where $t^m_{recom}$ is the time for recomputing $m$ KVs. With $M$ = 100K and $t^m_{recom}/m$ values 
obtained from Figure \ref{fig:KV_swap_recompute}, we can compute that $T_{break} \in [0.33\text{s}, 130\text{s}]$, with smaller values for larger $m$'s.
}


\kkim{While the KVs of longer requests (larger $m$) have smaller break-even intervals ($T_{break}$'s), it does not indicate that that they need to be evicted more quickly than the shorter requests. It does not tell \emph{which} requests to preempt but \emph{whether} it is cost-effective to preempt a request after a certain interval. If a KV is actually reused after its break-even interval, then it would have been more cost-effective to discard it from the cache and recompute later.
}

\begin{figure}[h!]
\centering
\includegraphics[width=0.55\columnwidth]{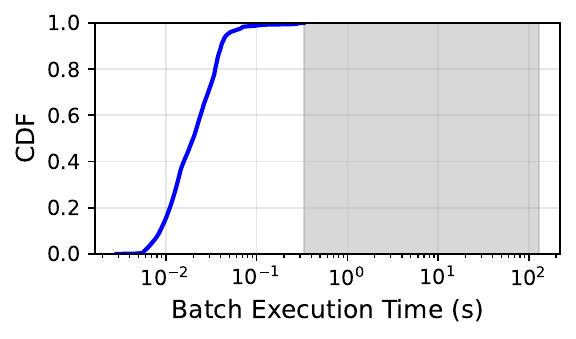}
\caption{\kkim{CDF of all batch execution times measured in Section \ref{subsec:sim_result:preemptive_simulation} for the KV cache size $M$ of 100K, where the batch times correspond to the KV reuse intervals. The gray region shows the range of break-even intervals.}}\label{fig:break_even_interval}
\end{figure}

\kkim{Figure \ref{fig:break_even_interval} shows the distribution of the reuse intervals for all workloads in Section \ref{subsec:sim_result:preemptive_simulation} and compares it with the break-even intervals. The batch times or TPOTs are reuse intervals as each batch re-accesses the stored KVs of running requests.
All reuse intervals are smaller than the break-even intervals, indicating that it is more cost-effective to keep all KVs in the cache.
This, however, shows that all KVs are important, and we need a different approach to select which requests to preempt.
}


\section{Theoretical Analysis and Shortest-Request First Replacement Policy}\label{sec:theory}







\kkim{This section provides theoretical analyses based on previous sections, starting from a global optimization problem for inference scheduling to deriving a new cache replacement policy for improving LLM inference performance. 
}

\subsection{Which Schedules are Optimal?}\label{subsec:theory:csp}


This section applies constraint satisfaction problem (CSP) \cite{CSP} to find optimal schedules \mlsys{and prove the insights found in the previous analysis in Section \ref{sec:sim_result}.} \mlsys{To the best of our knowledge, this is the first attempt to apply CSP to LLM inference scheduling. The closest work, \ExeGPT \cite{ExeGPT}, regards scheduler as a black box and searches for the best scheduling parameters out of the box, while we formulate the scheduling itself into CSP.
This is a more fine-grained approach, allowing higher flexibility and performance upper bounds, but with a trade-off of higher computational complexity.
As CSP utilizes the output lengths of requests, it is hypothetical. Unlike other schedulers that specify \emph{how} to schedule requests algorithmically, solving CSP gives \emph{what}: one or more schedules that 1) satisfy constraints to construct valid schedules and 2) achieve the best performance objectives, as well as the achievable performance bounds.}

Performance upper bounds allow for a more goal-oriented development of efficient inference techniques, determining promising ideas that are worth developing, unlike naively seeking a better scheduling algorithm without assured performance improvements, \kkim{or sweeping parameters to find the best system or scheduler configuration \cite{Vidur}.}
For instance, we can validate whether a better schedule exists that could reduce the latency of current schedules by 10\% for specific workloads, \kkim{and this schedule may not be generated from any existing schedulers.}
The optimization target can be adjusted to meet objectives such as latency, throughput, or fairness, if these can be represented in a formula.

In our CSP formulation, we first establish key notations. 
The index $i \geq 1$ and $j \geq 1$ represent the $i$-th request $r_i$ and $j$-th batch $\batch_{j}$ (each batch is a set of requests). We also use $j = 0$ as a virtual index to denote the initial system state, 
\kkim{$I_i$ and $O_i$ as input and output sizes of $r_i$,}
and $\I_{cond}$ as the indicator variable, which is 1 if a condition ${cond}$ is true and 0 otherwise.
\kkim{We use $C$ as the maximum number of tokens to process per batch and $M$ as the KV cache size.}
Now we explain the three parts of our CSP: variables, constraints, and objectives.

\noindent \underline{\textbf{Variables.}} 
\mlsys{$\sequence_{i,j}$ represents the number of all input and generated tokens for $r_{i}$ after processing $\batch_{j}$, with $\sequence_{i,0} = I_i$.
}
$\mem_{i,j}$ denotes \kkim{the number of KVs stored in the cache for $r_i$} \emph{after} processing $\batch_{j}$, with $\mem_{i,0} = 0$.
$c_{i,j}$ denotes \kkim{the number of tokens to process for $r_i$ at $\batch_{j}$,}
which is 0 if $r_i \not\in \batch_{j}$. 
$g_{i,j} = \I_{r_i \text{ generates a token at } \batch_{j}}$ and $e_{i,j} = \I_{r_i \text{ is preempted at } \batch_{j}}$.

\noindent \underline{\textbf{Constraints.}}
The constraints establish the interactions between variables and the conditions necessary for a valid schedule. 


\noindent \texttt{Termination:} Each request $r_i$ must generate $O_i$ output tokens, thus

\begin{equation}
\forall i: \sum_{j} \gen_{i,j} = O_{i}.
\end{equation}

\noindent \texttt{Per-Request Progress:} For the number of total input and generated tokens,

\begin{equation}
\forall i, j: \sequence_{i,j} = \sequence_{i,j-1} + g_{i,j}. 
\end{equation}

\noindent \texttt{Memory Management:} The number of KVs stored should be 0 if $r_i$ is preempted or increase by the number of processed tokens, $c_{i,j}$. 



\begin{equation}
\forall i, j: \mem_{i,j} =
\begin{cases} 
      0                       & \text{if } e_{i,j} = 1 \\
      \mem_{i,j-1} + c_{i,j}  & \text{otherwise}
\end{cases}
\label{eq:mem}
\end{equation}


\noindent \texttt{Tokens to Process:} The number of tokens to process should be 0 if $r_i$ is preempted or not exceed available tokens to process.


\begin{equation}
\forall i, j: c_{i,j} =
\begin{cases} 
      0                                     & \text{if } e_{i,j} = 1 \\
      \leq \sequence_{i,j-1} - \mem_{i,j-1} & \text{otherwise}
\end{cases}
\label{eq:c}
\end{equation}

\noindent \texttt{Token Generation:} The output token can only be generated if all input and generated tokens are processed, for both (chunked) prefill and decode steps.

\begin{equation}
\forall i, j: g_{i,j} =
\begin{cases} 
      1 & \text{if } c_{i,j} = \sequence_{i,j-1} - \mem_{i,j-1} \\
      0 & \text{otherwise}
\end{cases}
\label{eq:g}
\end{equation}

\noindent \texttt{Batch Constraints:} Ensures global constraints per batch.


\begin{equation}\label{eq:global}
\begin{split}
    \forall j: \sum_{i} c_{i,j} \leq C, \sum_{i} m_{i,j} \leq M
\end{split}
\end{equation}

The inequality in Equation (\ref{eq:c}) allows partial processing of the available tokens as in chunked prefill \cite{SARATHI}.
For implementing the conditional constraints in Equations (\ref{eq:mem})-(\ref{eq:g}), we use the Big-\textbf{M} method \cite{BigM} to linearize them, since linear programs are more efficient than non-linear ones \cite{LPandNLP}. For example, Equation (\ref{eq:mem}) is linearized as:
\begin{equation}
\begin{split}
\mem_{i,j} & \leq \text{\textbf{M}} \, (1 - e_{i,j}), \\
\mem_{i,j} & \leq \mem_{i,j-1} + c_{i,j} + \text{\textbf{M}} \, e_{i,j}, \\
\mem_{i,j} & \geq \mem_{i,j-1} + c_{i,j} - \text{\textbf{M}} \, e_{i,j},
\end{split}
\end{equation}
where \text{\textbf{M}} is a sufficiently large constant.
We implement our CSP using \Gurobi\footnote{https://www.gurobi.com}.

\noindent \underline{\textbf{Objective.}}
The objective can be set to minimize the total latency using our cost models in Section \ref{subsec:sim_design:cost_model}, \mlsys{summing up all batch times.} 
\todo{Since the cost models are monotonic, they prefer a lower number of batches and smaller number of tokens to process and KVs to read per batch.}
Instead of retrieving the minimum latency, one can simply opt for the existence of a better schedule, for example, by running a scheduler whose latency is $L$ and adding a constraint that the latency should be less than $0.9L$ to guarantee 10\% improvement.
Furthermore, one can optimize latency under throughput constraint, e.g., making TPOT or each batch time less than a predefined threshold. 


\noindent \underline{\textbf{Online Setting.}}
Supporting the online setting, where each request $r_i$ has an arrival time $T_i$, is straightforward: add a variable $Acc_{j}$ to track accumulated batch times and set $\sequence_{i,j} = m_{i,j} = 0$ if $Acc_{j} < T_{i}$ \kkim{to indicate that there is no token to process or processed before $T_{i}$.}


\noindent \underline{\textbf{Challenge.}}
A primary challenge in CSP is its limited scalability, as it is generally NP-complete \cite{CSPNP}. The complexity grows with the number of requests and batches, reaching millions of variables for thousands of requests and batches. Consequently, we primarily use CSP on small workloads that are enough to explain the phenomena found in the analysis and initialize the variables with outputs from other schedulers, rather than using all-zero or random values. We leave optimizing CSP for larger scales as a future work \mlsys{with possibly approximation approaches \cite{approximationforcsps, largescalecsp}.}


\subsection{\mlsys{Can Preemption Be Optimal?}}\label{subsec:theory:can_preemption_optimal}


As noted in Section \ref{subsec:sim_result:increasing_M}, preemption can enhance system performance. 
We use our CSP \kkim{first} as a \mlsys{proof-by-example approach.} We use four requests, set their output size $O$ to 4, and vary their input size $I$ from 1 to 1024, with the KV cache size $M$ of $max(2I, I+O-1)$.
This allows schedules to initiate prefills for two requests ($2I$) while ensuring that only one request can retain its KVs to generate the final token ($I+O-1 \leq M < 2(I+O-1)$). The objective is to minimize latency.

\begin{figure}[ht]
    \centering
    \subfigure[$I = 32$.]{%
        \includegraphics[width=0.23\textwidth]{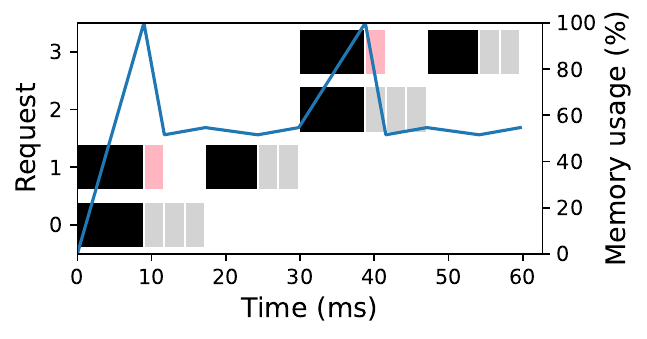}
        \label{fig:csp_I_32}
    }
    \hfill
    \subfigure[$I = 64$.]{%
        \includegraphics[width=0.23\textwidth]{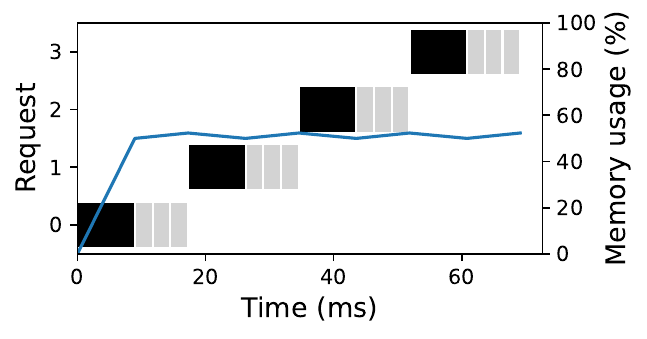}
        \label{fig:csp_I_64}
    }
    \vspace*{-0.2cm}
    \caption{CSP results for four requests, fixed output size $O$ of 4, and varying input size $I$. The KV cache size $M$ is $max(2I, I+O-1)$. Black, gray, and red boxes indicate requests in the (p)refill, decode, and preempt phases, respectively. Blue lines represent the KV cache usage.}
    \label{fig:csp_exp}
\end{figure}

Interestingly, the CSP also opts to preempt requests. 
For small input sizes ($I \leq 32$), CSP chooses to preempt short requests as illustrated in Figure \ref{fig:csp_I_32}, \todo{since it is faster to complete short prefills, making a progress of a token generation, and decreasing the number of batches by one, rather than waiting for another batch to retain memory. Each batch has an inevitable cost of loading model weights.}
This approach can reduce latency by \todo{up to 17\%} compared to preemption-free schedules as $I$ decreases.
However, for large input sizes ($I \geq 64$), CSP avoids preemption, as the later refill cost increases with the input size. Here, avoiding preemption can reduce latency by \todo{up to 40\%} as $I$ increases.
Among the schedulers, \vLLM and its preemption-free variant exhibit latency results close to the CSP's results, for $I \leq 32$ and $I \geq 64$, respectively.


\noindent \underline{\textbf{Remark.}}
\mlsys{While it is counterintuitive that preemption improves performance, we prove that it is true and find a relationship between sequence length and performance improvement by preemption.} 
Preempting short requests can help, but preempting long requests degrades performance due to high refill costs. 
\mlsys{Furthermore, Figure \ref{fig:csp_exp} shows that the cache is underutilized as in Figure \ref{fig:varying_M}, leaving an additional room for improvement.}

\subsection{\mlsys{\emph{When} Can Preemption Be Optimal?}}\label{subsec:theory:when_preemption_optimal}

\kkim{Due to the limited scalability of CSP, it is nearly infeasible to list all cases when preemption is optimal.
This section takes the insight from Section \ref{subsec:theory:can_preemption_optimal} and solely focuses on it, 
that when preempting and reprocessing requests later can reduce the number of batches, leading to reduced latencies.}

\kkim{We consider (1) processing $x$ tokens of a prefill request to generate an output token, (2) preempting the request, and (3) later refilling it. We can compute the increased overhead in this preemptive schedule compared to the corresponding non-preemptive schedule that does not process this request until there is enough space in the cache. First, for both schedules, (1) is an inevitable cost in processing the request. (2) adds no overhead to the preemptive schedule. (3) adds an overhead of processing $x+1$ tokens. Therefore, the increased cost is $F_{p}(x+1)$ where $F_p$
is the cost function for a prefill request in Section \ref{subsec:sim_design:cost_model}.
}

\kkim{Note that (3) also produces an output token, reducing one decode step for this request. Hence, we need to compute the saved cost from reducing one decode batch. This decode step in the non-preemptive schedule reads $x$ KVs from the memory and processes one token. 
The cost of this decode step is $F_d(x)$
where $F_d$ is the cost function for a decode request. 
}

\kkim{By comparing the two costs -- increased and saved -- in preemptive schedule, we can compute the range of $x$ that makes the preemptive one more efficient than the non-preemptive counterpart. This is equivalent to computing the range of $x$ such that $F_p(x+1) < F_d(x)$. 
This range is $x < 47$ for the setup in Section \ref{subsec:theory:can_preemption_optimal},
consistent with the results that preempting a request with input size $I = 32$ leads to a better efficiency.
}

\subsection{\mlsys{\emph{Which} Requests to Preempt?}}\label{subsec:theory:which_request_preempt}


\kkim{The previous section focused on processing and preempting the same request, identifying when it is helpful to preempt a request. This section targets \emph{which} requests to preempt if we have multiple candidates.}

\kkim{Based on the insight that preempting short requests can be helpful, we generalize this to prioritizing the preemption of short requests, while existing LLM inference systems simply preempt newest requests first, as shown in Table \ref{tab:llm-taxonomy}. Here, we prove that our \emph{shortest-request first} (SRF) policy is more efficient than their newest-request first (NRF) policy. SRF prioritizes running \emph{long} requests (having large number of KVs in the KV cache) and preempts \emph{short} requests if the KV cache reaches its limit.
}

\begin{figure}[h!]
\centering
\includegraphics[width=0.87\columnwidth]{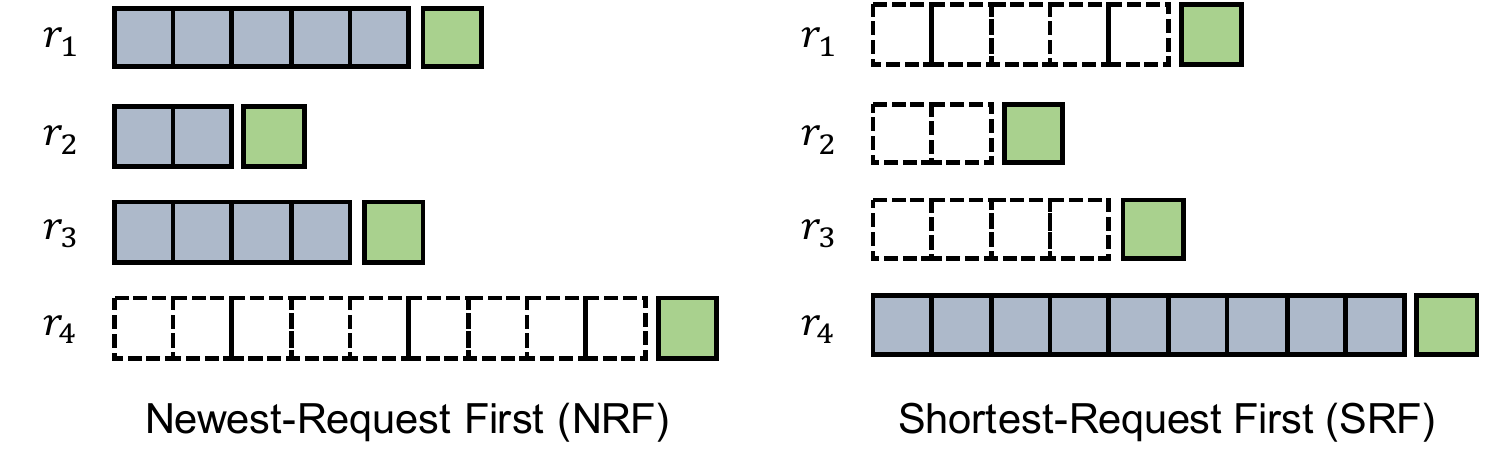}
\caption{\kkim{Example of NRF and SRF for four requests ordered by arrival times ($r_1$ to $r_4$) and freeing 9 KVs. Blue, empty, and green boxes represent stored KVs, evicted KVs, and unprocessed tokens.}}\label{fig:nrf_srf}
\end{figure}

\kkim{To determine which schedule is more efficient, we further abstract the cost-based analysis in Section \ref{subsec:theory:when_preemption_optimal} and introduce a concept called \emph{batch-wise progress}, which we define as the number of tokens generated over the batch execution time.
Since the batch time monotonically increases with the number of tokens (Section \ref{subsec:sim_design:cost_model}), we can simplify the progress as the number of tokens generated over the number of tokens to process.
Further, as the number of total tokens to generate is fixed across the schedules, maximizing the progress improves the  efficiency of a schedule.
If we preempt a request with $m$ KVs in the cache and $1$ output token generated but not processed, the progress of this request in the batch that refills this request, is $\frac{1}{m+1}$, from reprocessing $m$ KVs and $1$ last output token, and generating a new output token. Therefore, preempting a short request with small $m$ increases the progress, proving that the progress of using SRF is higher than using NRF. For the example in Figure \ref{fig:nrf_srf}, NRF's progress for refilling the newest request $r_4$ is $\frac{1}{9+1}$, while SRF's progress for refilling $r_1$, $r_2$, and $r_3$ are $\frac{1}{5+1}$, $\frac{1}{2+1}$, and $\frac{1}{4+1}$.
}


\section{Evaluation of SRF in Actual Inference Systems}\label{sec:deploy}











\kkim{This section evaluates our shortest-request first (SRF) preemption and cache replacement policy in Section \ref{subsec:theory:which_request_preempt} on actual inference systems and online workloads.}
\nexttodo{From Table \ref{tab:llm-taxonomy}, we choose three systems, \vLLM for the main system, \SGLang for prefix-sharing scenarios, 
and \Dynamo for disaggregation and KV offloading to DRAM. The first two are popular and open-sourced, implement all stacks in Figure \ref{fig:layers}, and often used as backends of other systems, including the last one.}

SRF is easy to implement in inference systems with just a few lines of code (order requests by the number of KVs in the cache) without any fine-tuning or training an ML model, not even our cost models (but still being cost-aware considering the progress defined).
At the point of cache insertion, we further maintain an optional, online histogram to estimate the output lengths of requests given their input lengths, predict if any preemption would occur for long-output requests, and defer scheduling those requests to later batches. We call this SRF+Hist.

\kkim{Table \ref{table:deploy_workloads} shows the workloads we used. \Azure \cite{AzurePublicDataset} is an 1-hour online trace of conversations, \LongForm \cite{LongForm} is a long-form text generation dataset to generate human-like texts, and \SelfConsistency \cite{SelfConsistency} is a prefix-sharing workload where each prompt generates multiple answers.
\ReasonTwo and \ChatFift are production traces provided from Alibaba \cite{servegen}.
}

\todo{Since the request lengths exceed the context size of Llama-2-7B, here we use Llama-3-8B and Llama-3-70B with the context size of 128K and set the KV cache size to 100K as before. This also shows the generalizability of our methodologies to other models.
}

\begin{table}[]
\renewcommand{\tabcolsep}{1mm}
\caption{\kkim{Workloads used to evaluate our SRF policy. NR is the number of requests, $I$ and $O$ denote request input and output sizes.}}
\label{table:deploy_workloads}
\begin{tabular}{c|ccccc}
\toprule
Workload           & NR & Avg. $I$ & Max. $I$ & Avg. $O$ & Max. $O$  \\ \hline
\Azure & 19.7K & 1.2K & 14.1K & 0.2K & 1K \\ 
\LongForm & 2K    & 0.3K  & 8.4K & 0.4K & 3.8K \\
\SelfConsistency & 1K & 0.3K  & 4.6K & 5.2K & 21K \\
\ReasonTwo & 1K & 0.9K  & 22.6K & 1.5K & 23.4K \\
\ChatFift & 4.4k & 0.8K  & 22K & 0.2K & 3K \\
\bottomrule
\end{tabular}
\end{table}

\subsection{Evaluation in \vLLM}\label{subsec:deploy:vllm}

\begin{figure}[h!]
\centering
\includegraphics[width=1.05\columnwidth]{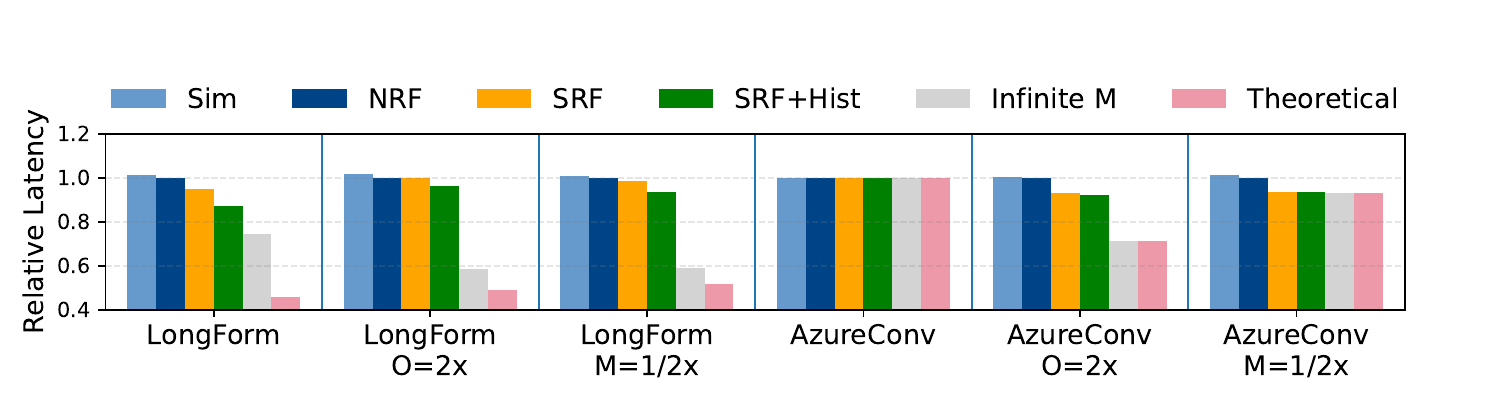}
\vspace*{-0.5cm}
\caption{\todo{Relative latencies of running inference workloads with different policies compared to the NRF in \vLLM, using Llama-3-8B on 1 A100. 
SRF and SRF+Hist are deployed in \vLLM. The others are simulated: \emph{Sim} simulates NRF, \emph{Infinite M} simulates an infinite-size KV cache, and \emph{Theoretical} simulates the hardware performance bounds in Equation (\ref{eq:latency_flops_data}).}}\label{fig:deploy_A100_8B}
\end{figure}

\kkim{This section presents the results on \vLLM, the main system we have used from Section \ref{sec:overview} and compared with the simulation results in Section \ref{sec:sim_result}.
SRF and SRF+Hist each required 6 and 24 lines of Python code to implement.}
\todo{Figure \ref{fig:deploy_A100_8B} shows the results for \kkim{\vLLM, using Llama-3-8B on 1 A100.} 
We also use the output length $O$ scale of 2x and KV cache size $M$ scale of 1/2x to evaluate on longer-generation and higher-contention scenarios. The results for \emph{Sim} shows that our simulation is accurate enough \xxx{(up to 2\% difference)}. Compared to NRF, SRF and SRF+Hist improve performance by \xxx{up to 8\% and 15\%} without performance regression. Note that these numbers are critical, indicating substantial savings in GPU hours and operational costs for high-demanding LLM inferences as mentioned in Section \ref{sec:introduction}. 
For our offline simulation workloads in Section \ref{sec:sim_result}, SRF improves performance by up to 40\% without regression due to higher arrival rates (all requests arrive at time zero) and higher contentions.
}




The last two policies in Figure \ref{fig:deploy_A100_8B} represent performance upper bounds, each for enhanced memory capacity and bandwidth utilization. While an infinite KV cache would save the latency up to 40\%, maximum bandwidth utilization would save up to 55\%. 
Furthermore, note that for \Azure, both NRF and SRF have similar latencies to the performance bounds, which is due to the low arrival rate that requests are processed without much contention. This highlights that simulation-based analysis allows identifying the workloads to optimize further or not, without spending extensive GPU hours.

\begin{figure}[h!]
\centering
\includegraphics[width=0.92\columnwidth]{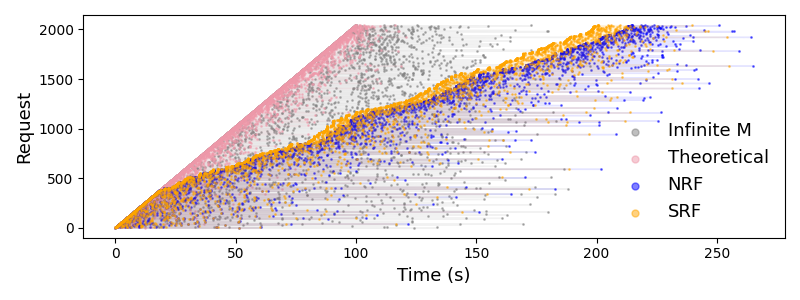}
\vspace*{-0.4cm}
\caption{\todo{\LongForm schedules visualized with start and completion times per request, under different policies.}}\label{fig:schedule_visualized}
\vspace*{-0.5cm}
\end{figure}

While increasing memory capacity and request concurrency reduces latency, it incurs an undesirable side effect. Figure \ref{fig:schedule_visualized} shows the schedules on \LongForm, where for \emph{Infinite M}, batching a large number of requests together leads to heavy batches as seen in Section \ref{sec:sim_result} and large request completion times. 
The figure also shows that SRF preserves fairness as in NRF, completing early-arrived requests first. This is because, SRF is a \emph{preemption and cache replacement} policy, where we still \emph{launch} earliest requests first.

\begin{figure}[h!]
\centering
\includegraphics[width=1.05\columnwidth]{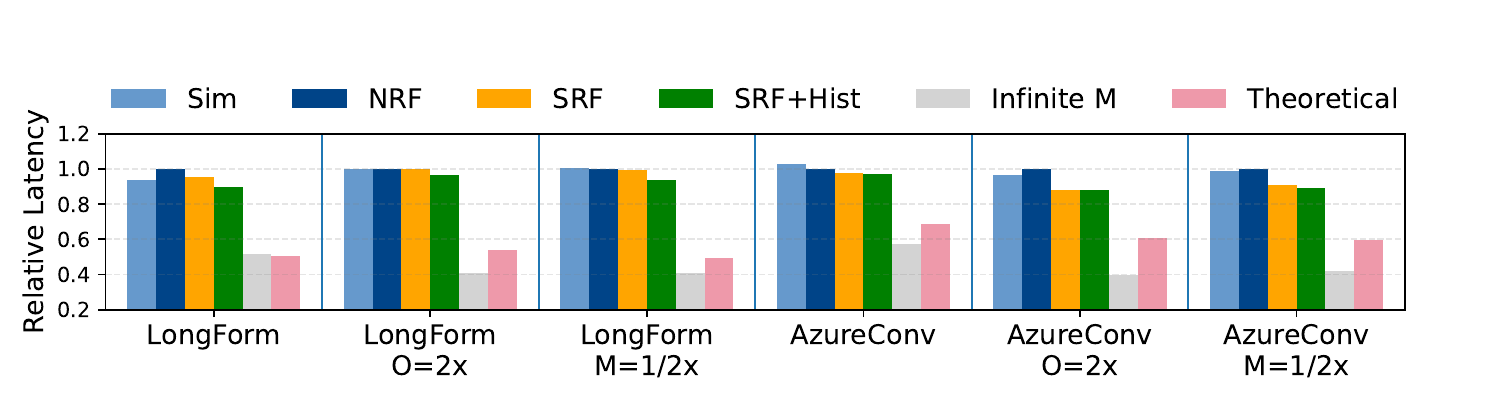}
\vspace*{-0.5cm}
\caption{\todo{Relative latencies, using Llama-3-70B on 4 A100.}}\label{fig:deploy_A100_70B}
\end{figure}

\kkim{Figure \ref{fig:deploy_A100_70B} shows the results for Llama-3-70B on 4 A100.} 
Both SRF and SRF+Hist improve performance by up to 13\% without regression. 
For \emph{Infinite M} and \emph{Theoretical}, increasing the memory capacity results in a larger performance improvement than increasing the bandwidth utilization, which is an opposite result from \kkim{Figure \ref{fig:deploy_A100_8B}}. This is due to the higher batch latencies that increase the contention level, where the arrived requests are not processed as promptly as in the 8B model, making the KV cache as the critical bottleneck. 

When we switched from A100 to H100, the contentions were largely resolved for the 8B model due to larger memory bandwidths, where SRF showed similar performance with NRF. 
For the 70B model, due to a higher contention level, SRF and SRF+Hist improved performance by up to 6\% and 11\%, and again, without performance regression.
\kkim{We omit the figures due to space limitations.}


\noindent \underline{\textbf{Remark.}} \kkim{SRF is easily deployable to an actual inference system and}
shows better performance than NRF without regression or losing fairness. 
\kkim{As shown in Section \ref{sec:sim_result}, GPU bandwidth is a primary factor affecting performance, and increasing it can reduce batch times. Under high contention due to large batch times, the capacity also matters.}


\subsection{Evaluation in \SGLang}\label{subsec:deploy:sglang}


\kkim{We compared NRF vs. SRF in \SGLang, a popular system optimized for prefix-sharing scenarios.
SRF and SRF+Hist each required 7 and 27 lines of Python code to implement.
We used the \SelfConsistency workload in Table \ref{table:deploy_workloads} with prefix-shared requests. \SGLang maintains a prefix tree of requests and preempts non-prefixes (leaf nodes in the tree, owned by a single request) first. If a non-prefix (and the owning request) is preempted, its parent node may become a new non-prefix which can be preempted. SRF sorts non-prefixes w.r.t. their lengths and preempts the shortest ones first.}

\kkim{SRF reduced the latency by 12\% and 16\% compared to NRF for Llama-3-8B on A100 and H100, and by 14\% and 17\% for Llama-3-70B on 4 A100 and 4 H100, respectively. SRF+Hist showed similar performance with SRF, and again, without performance regression compared to NRF.
}

\noindent \underline{\textbf{Remark.}} \kkim{SRF is easily deployable to prefix-sharing scenarios and improves performance there. The larger performance gain of SRF compared to Section \ref{subsec:deploy:vllm} is mainly due to the increased request output sizes of the workload that lead to higher contentions.}


\subsection{Evaluation in  \Dynamo}\label{subsec:deploy:dynamo}



\kkim{
We also deployed our SRF to \Dynamo under the prefill-decode disaggregation setup, using Llama-3-8B on two prefill and two decode A100.
SRF preempts requests in the decode side, and again, required less than 10 lines to implement.
We also enabled KV offloading to DRAM using \LMCache \cite{LMCache}.}





\kkim{
We used two production traces, \ReasonTwo and \ChatFift from Table \ref{table:deploy_workloads}, where the numbers indicate request-per-second. SRF improved performance by 7\% and 20\%, respectively.}

\noindent \underline{\textbf{Remark.}} \kkim{SRF improves performance in a production-scale, disaggregated system without modifying other stacks (e.g., request scheduling, disaggregation, or offloading policy).
Only a few lines in the preemption rule are changed, demonstrating the same lightweight integration and practical benefits highlighted from Section \ref{subsec:deploy:vllm}.}

\section{Conclusion}\label{sec:conclusion}

\kkim{In this work, we show how we can save GPU hours spent in both pre-deployment (development and analyis) and post-deployment (online inference serving) phases in LLM inference systems. We show that cost-based simulations are effective enough to provide meaningful insights and theoretical analyses, including that harnessing request preemptions can save GPU hours. We translate these into an effective preemption and cache replacement policy that can be easily deployed to actual systems and substantially save GPU hours for online workloads, which we expect greater benefits from longer-output requests with even higher contentions.
Considering the significant amount of inference usage, and as we target general inference workloads not specific to data analytics, we believe our work has a large impact to the community. 
We note that our work was inspired by the cost models, simple yet effective buffer management policies such as LRU, and the five-minute rule in DBMSs, and provides interesting connections between the database and LLM worlds. 
}
\bibliographystyle{spphys}       
\bibliography{sample_no_doi_url}

\end{document}